\documentclass[pre,twocolumn,floatfix,showpacs,amsmath,amssymb]{revtex4}
\usepackage{graphicx}
\usepackage{gensymb} 
\usepackage{epsfig}
\usepackage{multirow}
\begin{document}
\title{Strain-Rate Frequency Superposition in Large-Amplitude Oscillatory Shear}
\author{Chirag Kalelkar}
\email{kalelkar@gmail.com}
\author{Ashish Lele}
\email{ak.lele@ncl.res.in}
\author{Samruddhi Kamble}
\email{sb.kamble@ncl.res.in}
\affiliation{Complex Fluids and Polymer Engineering Group, Polymer Science and Engineering Division, National Chemical Laboratory, Pune 411008, India.}
\begin{abstract}
In a recent work, Wyss, {\it et.al.} [Phys. Rev. Lett., {\bf 98}, 238303 (2007)] have noted a property of `soft solids' under oscillatory shear, the so-called strain-rate frequency superposition (SRFS). We extend this study to the case of soft solids under large-amplitude oscillatory shear (LAOS). We show results from LAOS studies in a monodisperse hydrogel suspension, an aqueous gel, and a biopolymer suspension, and show that constant strain-rate frequency sweep measurements with soft solids can be superimposed onto master curves for higher harmonic moduli, with the {\it same} shift factors as for the linear viscoelastic moduli. We show 
that the behavior of higher harmonic moduli at low frequencies in constant strain-rate frequency sweep measurements is similar to that at large strain amplitudes in strain-amplitude sweep tests. We show surface plots of the harmonic moduli and the energy dissipation rate per unit volume in LAOS for soft solids, and show experimentally that the energy dissipated per unit volume depends on the first harmonic loss modulus alone, in both the linear and the nonlinear viscoelastic regime.
\end{abstract}
\maketitle
\section{Introduction}
Linear viscoelasticity\cite{Ferry} is limited to small strains or strain rates, and assumes a constant dynamic viscosity and zero normal stress differences in shear flows. In order to probe their linear viscoelastic response, materials are frequently subject to {\it small}-amplitude oscillatory shear, wherein an applied sinusoidal strain $\gamma(t)=\gamma_0 sin(\omega t)$ (here $\gamma_0$ is the strain amplitude, $\omega$ is the angular frequency, $t$ is the time) results in a stress response $\sigma(t;\omega)=\gamma_0[G'_1(\omega)sin(\omega t)+G''_1(\omega)cos(\omega t)]$ at the same angular frequency as the input. Here $G'_1$ and $G''_1$ are the so-called `storage' and `loss' modulus\cite{Ferry}, and are functions of the angular frequency. By contrast, the resultant stress from a {\it large}-amplitude oscillatory shear (LAOS)\cite{Philippoff,Onogi,Krieger,Tee,Davis,Ganeriwala,Gamota,Reimers,Collyer,Ewoldt} test contains higher harmonics which may be interpreted in terms of harmonic moduli $G'_n$ and $G''_n$ (subscript $n$ refers to the $n$-th harmonic, see Section III(A) for the definition), which are functions of both the strain amplitude and the angular frequency. The moduli $G''_1$ may be physically interpreted\cite{Ganeriwala} in terms of the energy dissipated per unit volume per cycle of strain oscillation (see Section III(E) below). In a recent work, Ewoldt, {\it et.al.}\cite{Ewoldt} have assigned a physical meaning to the third harmonic moduli in terms of a deviation from linear viscoelastic behavior by decomposing the stress response into a superposition of an elastic and a viscous component which are then expanded in terms of Chebyshev polynomials of the first kind. The signs of the third-order Chebyshev coefficients have been interpreted in terms of strain stiffening/softening of the material using the elastic stress component, and in terms of shear thickening/thinning using the viscous stress component. Meissner\cite{Meissner} carried out a comprehensive rheological study of three low-density polyethylene samples which were practically indistinguishable through usual characterization methods, and in their linear viscoelastic response, but behaved differently when subject to non-linear deformation processes such as film blowing and extrudate swell. Apart from their obvious utility in industrial processes, LAOS studies are of fundamental interest in rheology, and may lead to the development of more representative constitutive models.

In a recent work, Wyss, {\it et.al.}\cite{Wyss}, have shown a suprising feature in `soft solids' (compressed emulsions, concentrated suspensions, foams, pastes, gels, {\it etc.}), wherein constant strain-rate frequency sweep (SRFS) measurements of $G'_1$ and $G''_1$ may be superimposed onto master curves, with the structural relaxation time $\tau(\dot{\gamma_0})$ of the material showing an inverse power-law dependence\cite{Leonardo} on the strain-rate amplitude $\dot{\gamma_0}$ (at high strain-rate amplitudes), viz. 
$\tau({\dot{\gamma_0}})\propto\dot{\gamma_0}^{-\nu}$ ($\nu>0$ is the exponent in the power-law). In typical soft solids, the linear viscoelastic moduli cross over at a frequency which is too low to be probed via conventional rheological tests, and the utility of the SRFS procedure is that by increasing the strain-rate amplitude, the crossover frequency can be shifted to a range where it may be directly probed. Apart from this, the SRFS procedure supplies a unified picture of the slow relaxation dynamics\cite{Sollich} in a wide range of soft materials and is of intrinsic interest. Similar results have been confirmed in a micelle-forming triblock copolymer\cite{Mohan}, aqueous foams\cite{Marze}, and dough mixed with gluten\cite{Desai}. However, the results in Wyss, {\it et.al.}\cite{Wyss} have only been interpreted in terms of {\it linear} viscoelastic moduli, although the authors have carried out several tests with strain amplitudes in the non-linear viscoelastic regime of their samples.

In this paper, we present results from a systematic LAOS study of three representative soft solids, and show that higher harmonic moduli in SRFS measurements can also be rescaled onto master curves, with the {\it same} shift factors as for the linear viscoelastic moduli. We show that the higher harmonic moduli in SRFS measurements at low frequencies are similar to the moduli calculated at large strain amplitudes in strain-amplitude sweep tests. We show surfaces of the harmonic moduli and the energy dissipation rate per unit volume in LAOS and explicitly verify an earlier theoretical result\cite{Ganeriwala} that the energy dissipated per unit volume in a viscoelastic material depends on the first harmonic loss modulus alone in both the linear and the nonlinear viscoelastic regime. In addition, we supply an unambiguous and detailed prescription for carrying out LAOS studies with soft solids.
\section{Experiment}
\subsection{Instrumentation}
Our experiments were carried out at room temperature on the strain-controlled Advanced Rheometric Expansion System-$2000$ (ARES-$2000$, TA Instruments, United States) rheometer.  We used a cone-plate measuring system\cite{cp} with a cone diameter of $25$ mm (cone angle=$0.1$ rad). The rheometer permits  acquisition of DC voltage signals from the torque transducer (for measurement of torque) and the optical encoder (for measurement of the motor angular deflection) through BNC connectors in the rear panel of the instrument. These unprocessed voltage signals (in the range $\pm5$ volts) are not noise-filtered, or corrected for inertia and compliance of the torque transducer. Data was acquired at $16$-bit analog input resolution through an analog-to-digital card (NI PCI-$6014$, National Instruments, United States) coupled with a Labview (National Instruments, United States) code at a sampling rate\cite{nyquist} of $10^3$ points per second. A small DC offset was subtracted from the acquired oscillatory signal, and the signal was filtered for noise using a Savitzky-Golay filter\cite{Press}. The signal was then calibrated\cite{autorange} to find formulae which were used to convert the voltage values to quantities of physical interest. The calibration curves used were $y=0.046x$ ($x$ in volts, $y$ in Nm) for the torque and $y=0.1x$ ($x$ in volts, $y$ in radians) for the deflection angle. The values of the stress (in Pascals) and the strain were calculated from the torque and the deflection angle respectively using conversion factors appropriate to the measuring-system geometry and torque transducer employed. 
\subsection{Material synthesis and characterization}
The materials used in our tests were Poly N-isopropylacrylamide (PNIPAM), Xanthan gum, and Brylcreem Wetlook Gel. After loading the sample on to the rheometer, a thin layer of silicone oil (SF 1000, GE Bayer Silicones, India) was applied at the edges of the sample to prevent drying.

PNIPAM synthesis was carried out in a double-jacket glass kettle reactor by a free-radical polymerization reaction\cite{Senff}. The temperature was controlled by a water-circulating bath (Cool Tech $320$, Thermo-Electron Corp., United States). $600$ ml of distilled, deionised `Millipore water' (Milli-Q Gradient A$10$, Millipore, United States) purged with nitrogen gas for $1$ hour, was used for preparing the solutions. The monomer NIPAM (Acros Chemicals, Belgium), the cross-linker N,N'-methylene-bisacrylamide BIS (Sigma-Aldrich, United States) and the initiator potassium peroxodisulfate KPS (Sigma-Aldrich, United States) were recrystallized from appropriate solvents, and vacuum dried at room temperature for $4$ hours. $7.87$g of NIPAM, $0.39$g of BIS, and $0.15$g of the stabiliser sodium dodecyl sulfate (Merck, United States) were mixed in $480$ml of water at $25\celsius$, and stirred using an overhead stirrer (RZR $2051$control, Heidolph Instruments, Germany) at $100$ rpm for $30$ minutes under an inert atmosphere. $0.3$g of KPS in $20$ ml Millipore water, was added to the reaction mixture at $70\celsius$, and stirred at $300$ rpm, the reaction being allowed to proceed for $4$ hours. The temperature was then reduced to $25\celsius$, and the reaction mixture stirred overnight at $100$ rpm. Finally, the dispersion was dialysed (using dialysis bags having a molecular weight cutoff of $10000$ g/mol) against Millipore water for two weeks. The dialysed sample was frozen using liquid nitrogen, freeze-dried in a freeze drier (Heto Power Dry LL$3000$, Thermo-Electron Corp., United States) for $8$ hours and stored in a 
dessicator. The prepared aqueous PNIPAM dispersion had a concentration of $14.09\%$ w/w. The hydrodynamic radius of PNIPAM was found to be $63.9$ nm by Dynamic Light Scattering (DLS) on a Particle Size Analyser 
(BIC $90$Plus, Brookhaven Instrument Corp., United States).

Xanthan gum ($4\%$ w/w) was prepared by dissolving $4$g of Xanthan gum (Sigma-Aldrich, United States) 
in $100$g Millipore water, under constant stirring using a Heidolph overhead stirrer for $8$ hours. $7$ mg 
of sodium azide (Merck, United States) was added to prevent bacterial growth.

Brylcreem Wetlook Gel (Godrej Sara Lee Ltd., India) was purchased off-the-shelf and used as received.
\section{Results and Discussion}
\subsection{Large-amplitude oscillatory shear (LAOS)}
Using results from continuum mechanics, Christensen\cite{Christensen} has shown that for an imposed sinusoidal strain $\gamma(t)=\gamma_0 sin\hspace{0.1cm}\omega t$ (here $\gamma_0$ is of arbitrary magnitude), the resultant stress in an isotropic material may be written as an odd-harmonic Fourier series:
\begin{eqnarray}
&\sigma(t;\omega,\gamma_0)=\gamma_0\sum_{n=1,3,5,...}[G'_n(\omega,\gamma_0)sin\hspace{0.1cm}n
\omega t\nonumber\\
&+G''_n(\omega,\gamma_0)cos\hspace{0.1cm}n\omega t],
\label{eq:memoryint}
\end{eqnarray}
where $G'_n$ and $G''_n$ are $n$-th harmonic moduli which are, in general, functions of both the 
angular frequency and the strain amplitude. 

We define the $n$-th harmonic moduli as 
\begin{eqnarray}
G'_n(\omega,\gamma_0)\equiv\frac{\sigma_n}{\gamma_1}cos\hspace{0.1cm}\Phi_n, \nonumber\\
G''_n(\omega,\gamma_0)\equiv\frac{\sigma_n}{\gamma_1}sin\hspace{0.1cm}\Phi_n, 
\label{eq:moduli}
\end{eqnarray}
where $\Phi_n(\omega,\gamma_0)=\phi_{n,\sigma}-n\phi_{\gamma}$ are the phase angles for the $n$-th harmonic moduli ($\phi_{n,\sigma}$ and $\phi_{\gamma}$ are the phase angles of the $n$-th harmonic in the stress and the first harmonic in the strain Fourier series, respectively). Here $\sigma_n(\omega,\gamma_0)$ is the $n$-th harmonic in the stress amplitude spectrum, and $\gamma_1$ is the first harmonic in the strain amplitude spectrum\cite{ampspec}. Note that in our definition of the moduli [Eqs. (\ref{eq:moduli})] we distinguish between $\gamma_0$, the {\it requested} strain amplitude, and $\gamma_1$, the {\it measured} amplitude of the first harmonic in the strain amplitude spectrum - these quantities differ by a small amount, which was empirically found to vary with the torque transducer compliance.
\subsection{Data analysis}
The general stress response in LAOS [Eq. (\ref{eq:memoryint})] may be rewritten as $\sigma(t;\omega,\gamma_0)=\sum_{n=1,3,5,...}\tilde{\sigma}_n sin(n\omega t+\phi_{n,\sigma}$), where $\tilde{\sigma}_n$ and $\phi_{n,\sigma}$ are both real-valued functions of $\omega$ and $\gamma_0$. In LAOS theory, the strain is assumed to be a sine wave with zero phase, but an experimental strain signal [$\gamma(t)=\gamma_0sin(\omega t+\phi_{\gamma}$)] has a non-zero phase angle $\phi_{\gamma}$ whose value depends on the instant of time when data acquisition commences. Hence, it is useful to apply the following transformation: $t=t'-\phi_{\gamma}/\omega$, and obtain $\gamma(t')=\gamma_0sin(\omega t')$ and $\sigma(t';\omega,\gamma_0)=\sum_{n=1,3,5,...}\tilde{\sigma}_n sin(n\omega t'+\phi_{n,\sigma}-n\phi_{\gamma}$). Therefore, data analysis must be carried out on complete oscillation cycles of the strain signal, with zero phase.

Data analysis was carried out using Matlab (The MathWorks Inc., United States). The full-width at half maximum (FWHM) of the sinc function\cite{sinc} is approximately $1/(2T)$\cite{Jyrki}, where $T$ is the duration of the time series. We choose the duration of our runs to correspond to a value of FWHM equal to $0.5\%$ of the imposed angular frequency, this corresponds to a run with at least $50/\pi$ cycles of oscillations. In the ARES-$2000$ rheometer, the measured strain amplitude in oscillatory shear, reaches the requested value $\gamma_0$ after a few transient cycles of oscillation. We ensure that our signal is processed from the instant of time when the measured strain amplitude is close to the requested value, and use a zero-crossing algorithm to extract $50$ cycles of oscillation from the strain and the stress signal (which have the same time-base). Note that for the materials under study, the amplitude of the measured oscillatory torque signal does not decay appreciably during the course of the experiments. We apply a discrete Fourier transform on the extracted signals, and use only the first half of the Fourier-transformed dataset (remainder of the dataset is redundant for real input data). Although in theory, the higher harmonic frequencies are integer multiples of the fundamental frequency, in an experimental stress signal the measured harmonic frequencies located through peak values in the stress amplitude spectrum, differ by a small amount from the integer multiples. Hence, we use a code which locates peak values in the stress amplitude spectrum and explicitly finds the frequencies corresponding to the harmonics. Finally, we calculate the phase angles and thereafter the harmonic moduli of interest.

In this study, we do not apply oversampling\cite{Dusschoten} techniques to improve the signal-to-noise ratio, and the resolution of the moduli beyond the fifth was found to be poor, due to noise and the low sampling rate. In addition, although the first harmonic moduli reported by the proprietary software accompanying the ARES-$2000$ rheometer are reliable up to $\omega=100$ rad/s (the upper limit for the machine), we found that the calculated higher harmonic moduli are of relatively low resolution beyond $\omega\approx50$ rad/s. Therefore, we restrict our study of the higher harmonic moduli up to the fifth order, with a maximum angular frequency of $50$ rad/s. It is important to note that there is no automated means of obtaining higher harmonic moduli from the ARES-$2000$ rheometer, therefore, each point in our graphs (with the exception of Figs. \ref{fig:ampfreqsweep_pnipam} and \ref{fig:spectra_pnipam}) is calculated from an {\it independent} oscillatory shear test with a fixed angular frequency and strain amplitude.
\subsection{Strain-rate frequency superposition (SRFS) in LAOS}
Most of the results that follow are from experiments carried out with our highly monodisperse PNIPAM sample, although a few results are also reported from studies with Xanthan gum. The results from tests carried out with Brylcreem are reported in one instance. 

\begin{figure}
\includegraphics[height=1.8in]{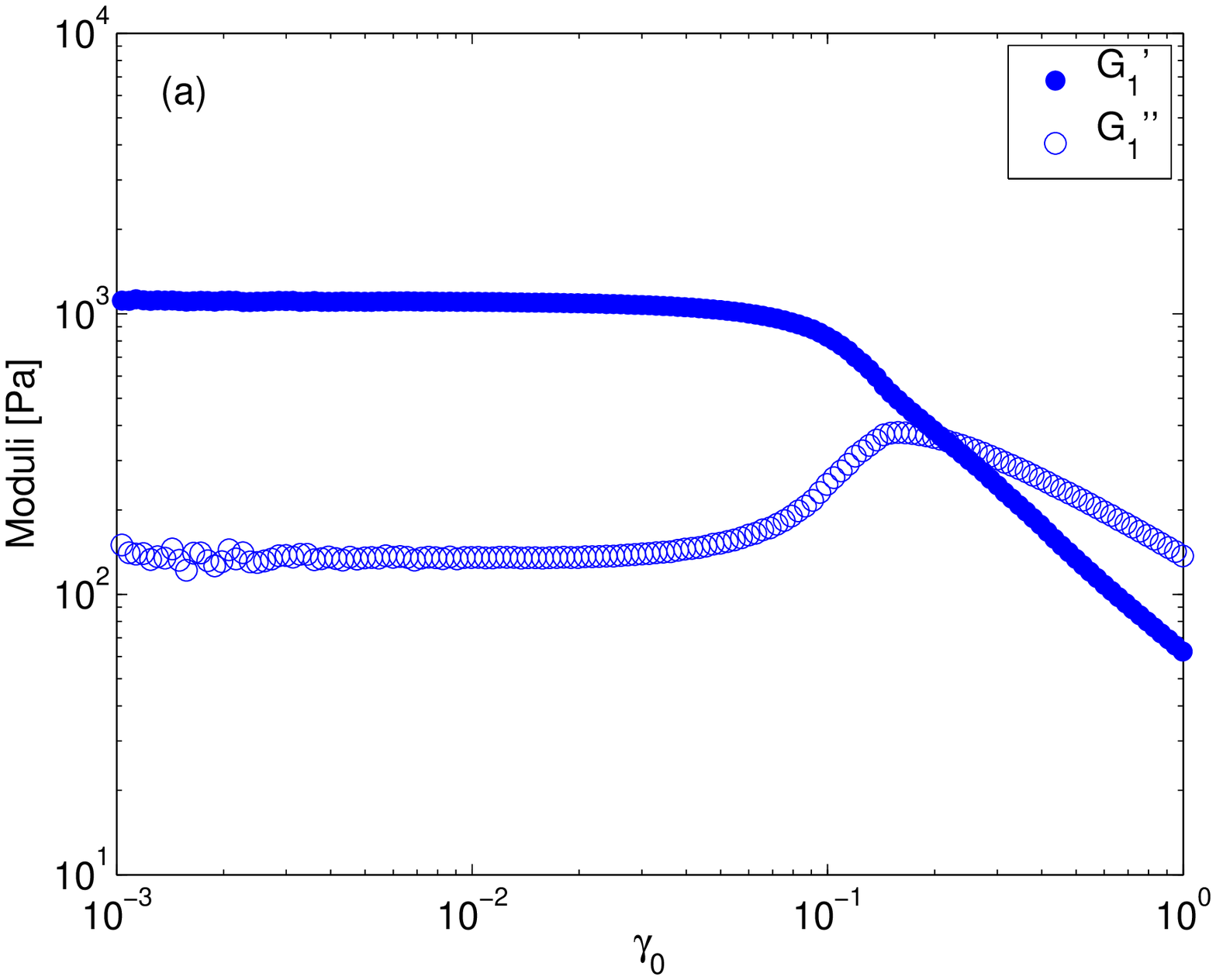}
\includegraphics[height=1.8in]{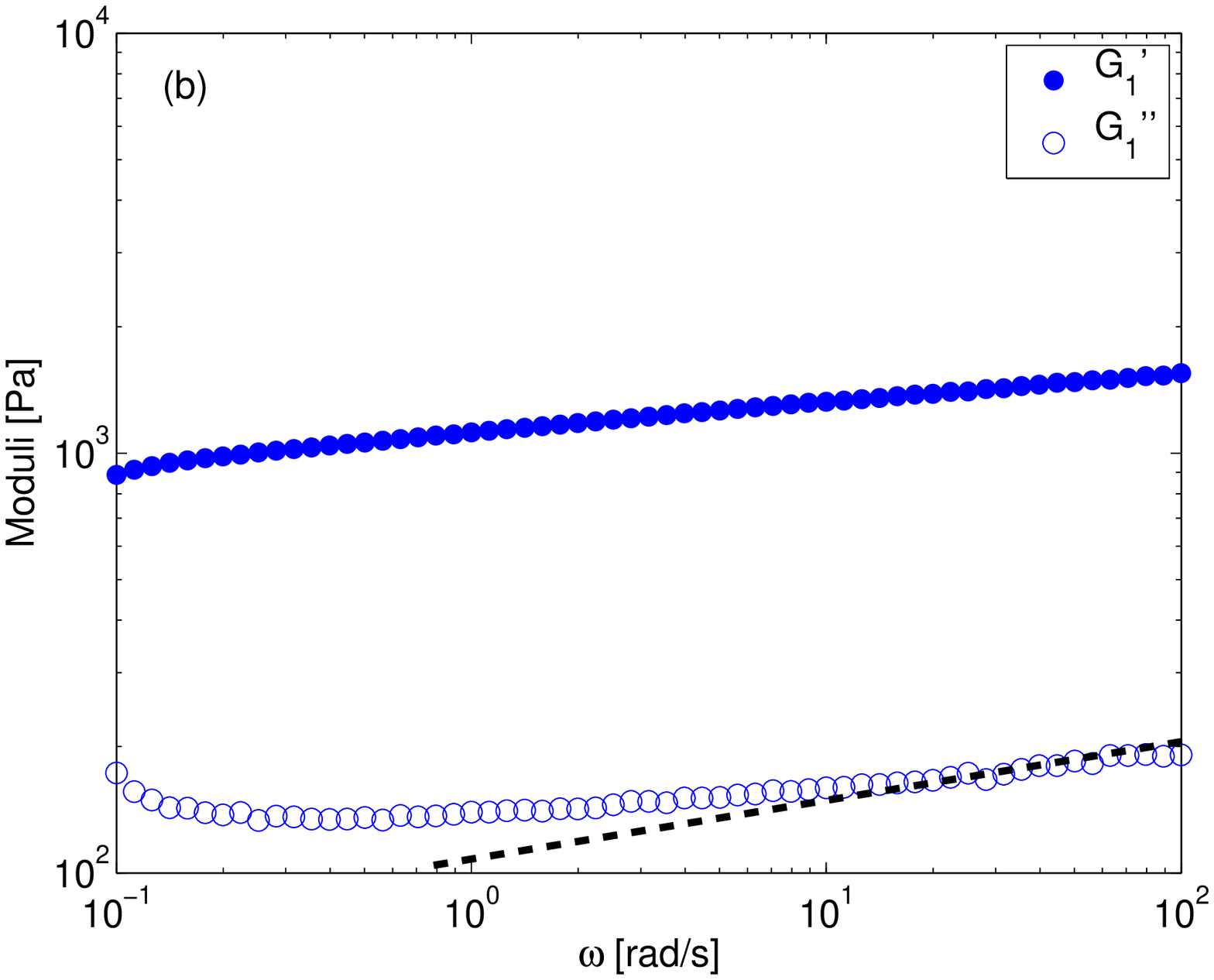}
\caption{\label{fig:ampfreqsweep_pnipam}(a) Plot of the first harmonic moduli $G'_1$ (solid circles) and 
$G''_1$ (open circles) as a function of varying strain amplitude $\gamma_0$  with $\omega=1$ rad/s using PNIPAM.\\(b) Plot of the first harmonic moduli $G'_1$ (solid circles) and $G''_1$ (open circles) as a function of varying angular frequency $\omega$ with $\gamma_0=0.01$ using PNIPAM. The dashed line is proportional to $\omega^{0.078}$ (see the text for discussion).} 
\end{figure}

\begin{figure}
\includegraphics[height=1.8in]{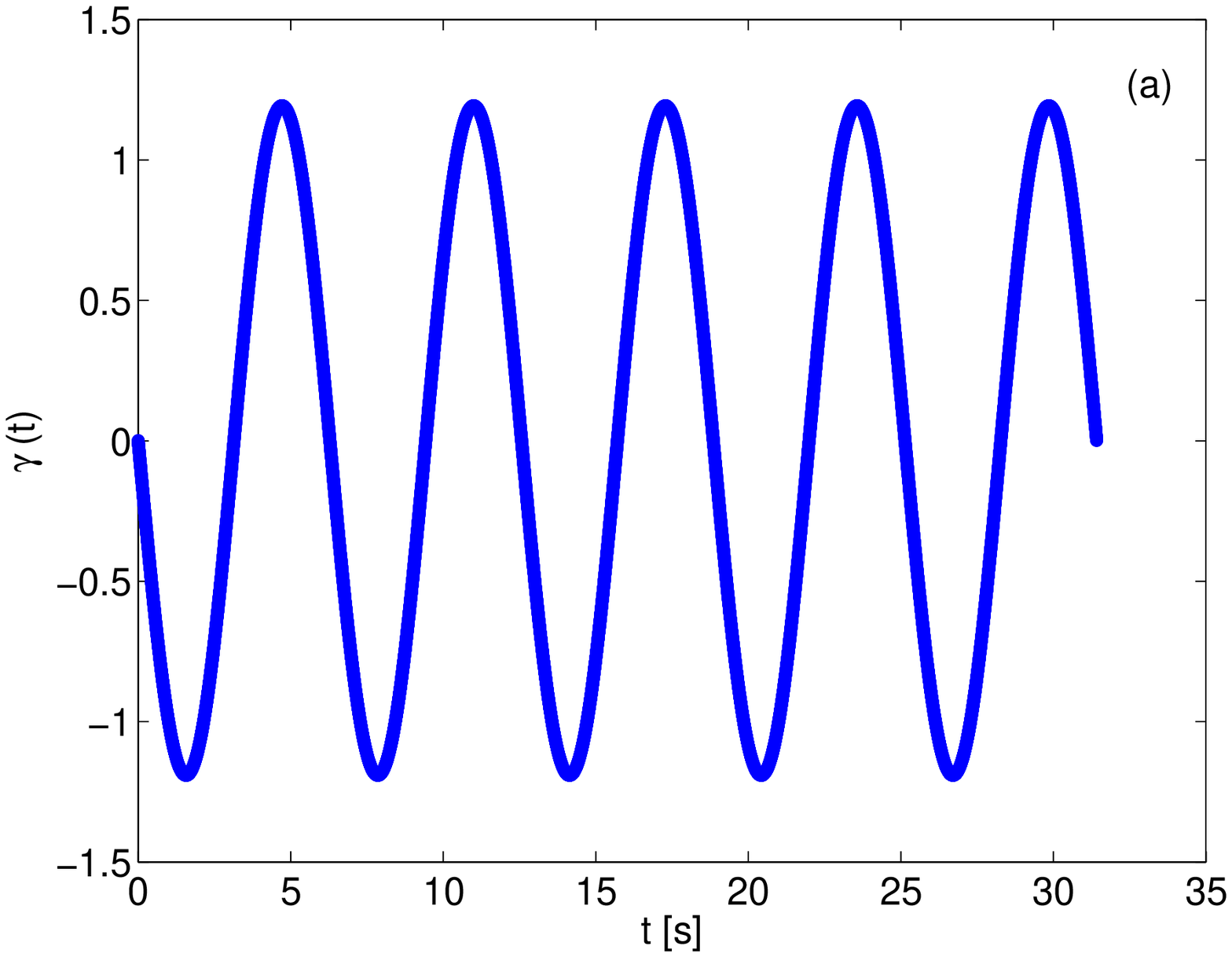}
\includegraphics[height=1.8in]{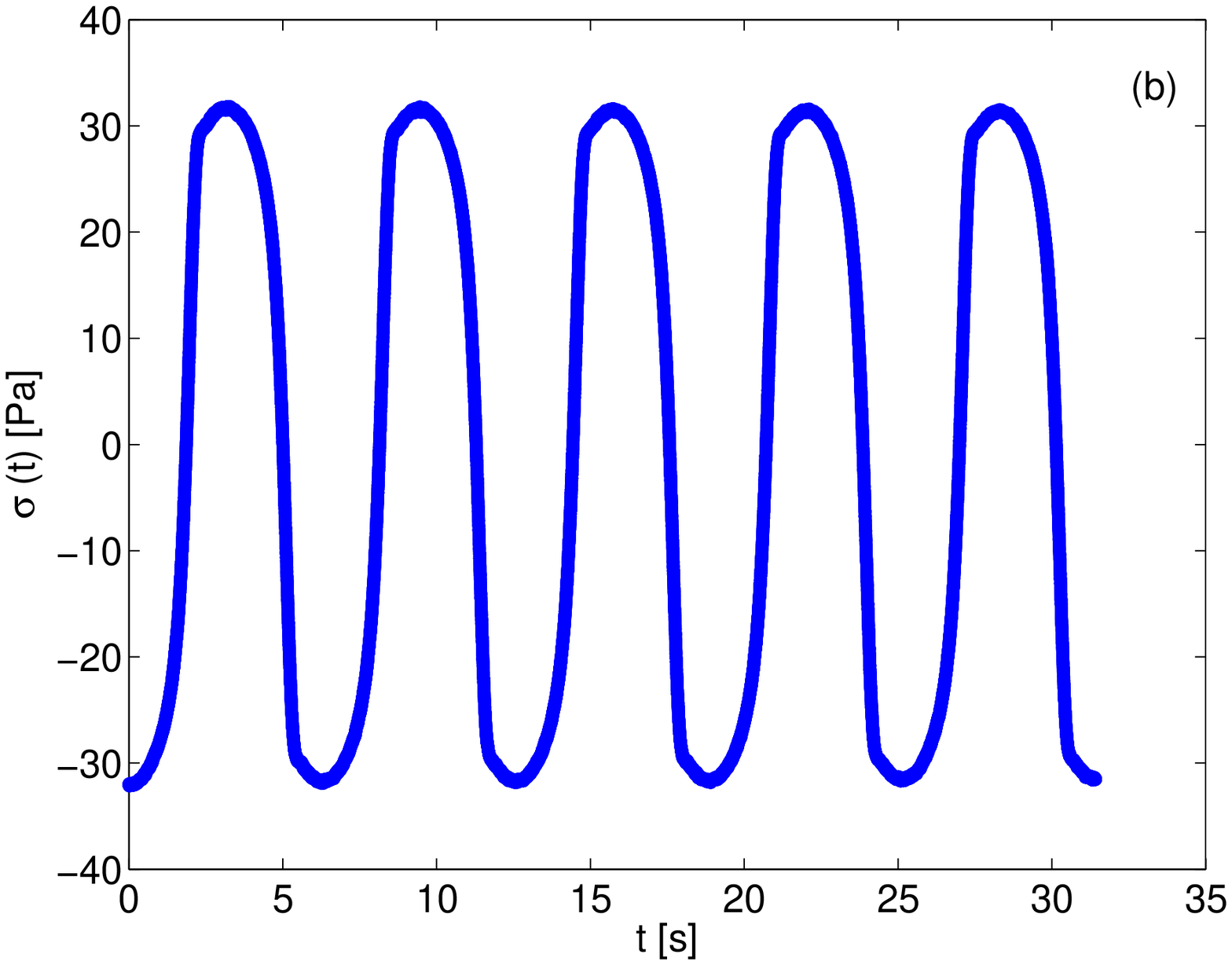}
\includegraphics[height=1.8in]{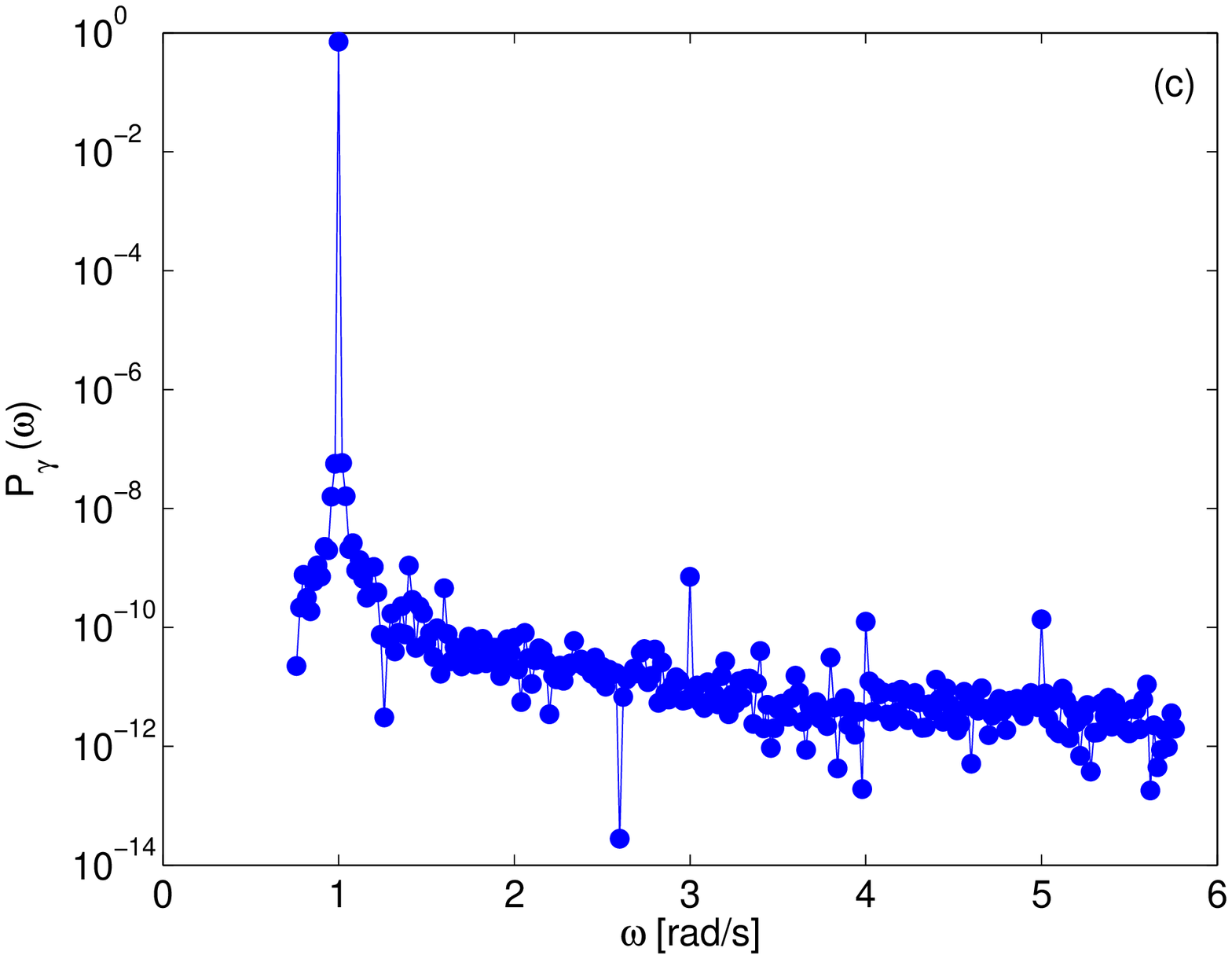}
\includegraphics[height=1.8in]{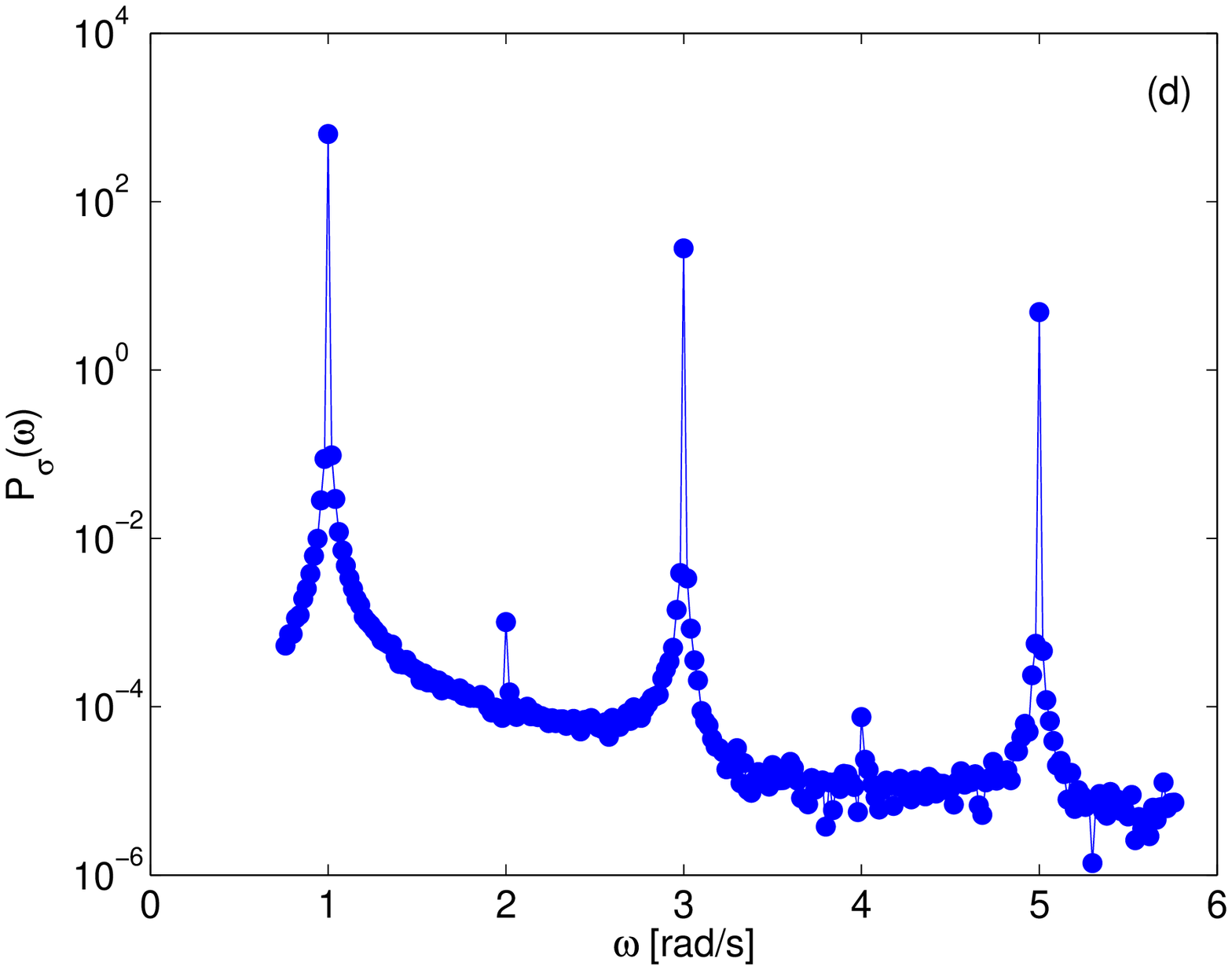}
\caption{\label{fig:spectra_pnipam}(a) Plot of a typical strain signal $\gamma(t)$ as a function of the time $t$ from an oscillatory shear test with $\gamma_0=1.2$, $\omega=1$ rad/s using PNIPAM.\\(b) Plot of the resultant stress signal $\sigma(t)$ as a function of the time $t$ with parameter values as in (a).\\(c) Plot of the power spectrum $P_\gamma$ (see the text for the definition) of the strain signal (a few oscillations are shown in (a)) as a function of the angular frequency $\omega$.\\(d) Plot of the power spectrum $P_\sigma$ of the stress signal (a few oscillations are shown in (b)) as a function of the angular frequency $\omega$.} 
\end{figure}

In Fig. \ref{fig:ampfreqsweep_pnipam}(a), we plot the first harmonic moduli $G'_1$ and $G''_1$ as a function of varying strain amplitude $\gamma_0$ with the angular frequency $\omega=1$ rad/s, using PNIPAM. The plot shows features characteristic of soft solids, with a pronounced peak in $G''_1$ at intermediate frequencies and a corresponding fall in $G'_1$, indicative of a breakdown of internal structure, and resultant liquid-like behavior at higher strain amplitudes\cite{Wyss}. $G''_1$ is a measure of the energy dissipated per unit volume per cycle of strain oscillation (see Section III(E) below) and the growth towards a peak value may be interpreted in terms of energy loss (as heat), on account of the work done to break internal structures in the material. In Fig. \ref{fig:ampfreqsweep_pnipam}(b), we plot the first harmonic moduli as a function of $\omega$ for $\gamma_0=0.01$, a strain amplitude within the linear viscoelastic regime of PNIPAM. In the range of frequencies probed, the sample shows solid-like behavior with $G'_1$ greater than $G''_1$, but the small upturn in $G''_1$ at the smallest frequencies hints at the possibility of a crossover of the moduli at a much lower frequency. The dashed-line curve in this figure is discussed below. \\
In Fig. \ref{fig:spectra_pnipam}(a), we plot a few cycles of sinusoidal oscillations of the strain signal as a function of the time from a LAOS test at $\gamma_0=1.2$, $\omega=1$ rad/s using PNIPAM. In Fig. \ref{fig:spectra_pnipam}(b), we plot the resultant stress signal as a function of the time. Note the non-sinusoidal nature of the stress signal, indicative of higher harmonics in the stress amplitude spectrum. In Fig. \ref{fig:spectra_pnipam}(c), we plot the strain power spectrum $P_\gamma$ as a function of $\omega$ for $50$ oscillation cycles of the strain signal, a few oscillation cycles of which are shown in Fig. \ref{fig:spectra_pnipam}(a). Here $P_h(\omega)\equiv2|H(\omega)|^2$ is the one-sided power spectrum of the real-valued time-series $h(t)$ with Fourier coefficients $H(\omega)$. The higher harmonics in the measured strain power spectrum may be considered as part of broadband noise. Our definition of the moduli [Eqs. (\ref{eq:moduli})] and resulting calculations do not account for higher harmonics in the strain signal. In Fig. \ref{fig:spectra_pnipam}(d), we plot the response stress power spectrum $P_\sigma$ as a function of $\omega$. Prominent odd harmonics\cite{nonlinmon} can be seen at frequencies which are close to integer multiples of the applied oscillation frequency $\omega=1$ rad/s along with much smaller even harmonics, a result first noted by Krieger and Liu\cite{Krieger}. Even harmonics in the stress power spectrum have been hypothesized\cite{Adrian} to occur in oscillatory shear flows showing wall-slip, or due to secondary flows in the plate gap. In our experiments, the ratio of the second to the first harmonic in the stress power spectrum is of order $10^{-6}$ or smaller, even harmonics are therefore neglected.

\begin{figure}
\includegraphics[height=1.8in]{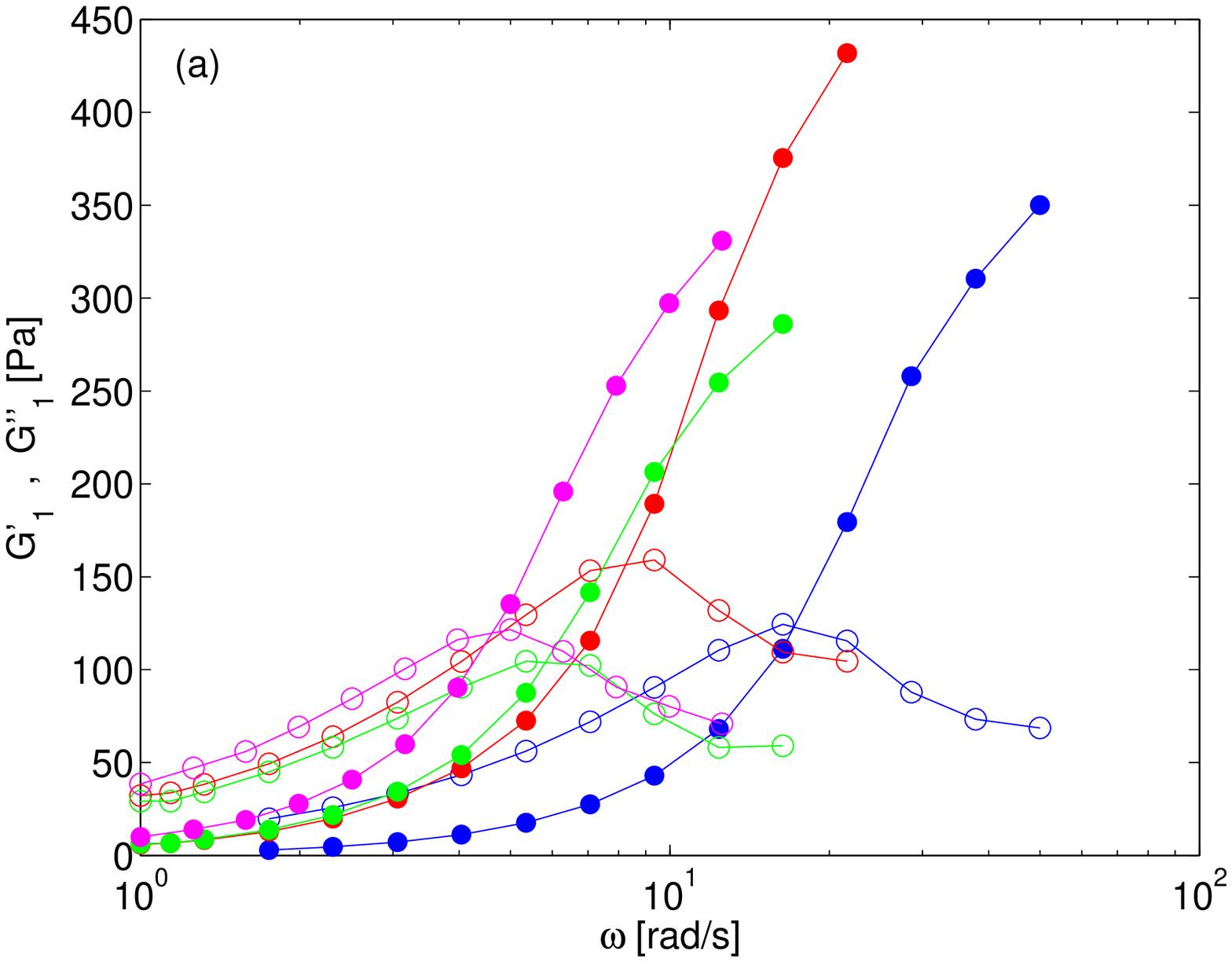}
\includegraphics[height=1.8in]{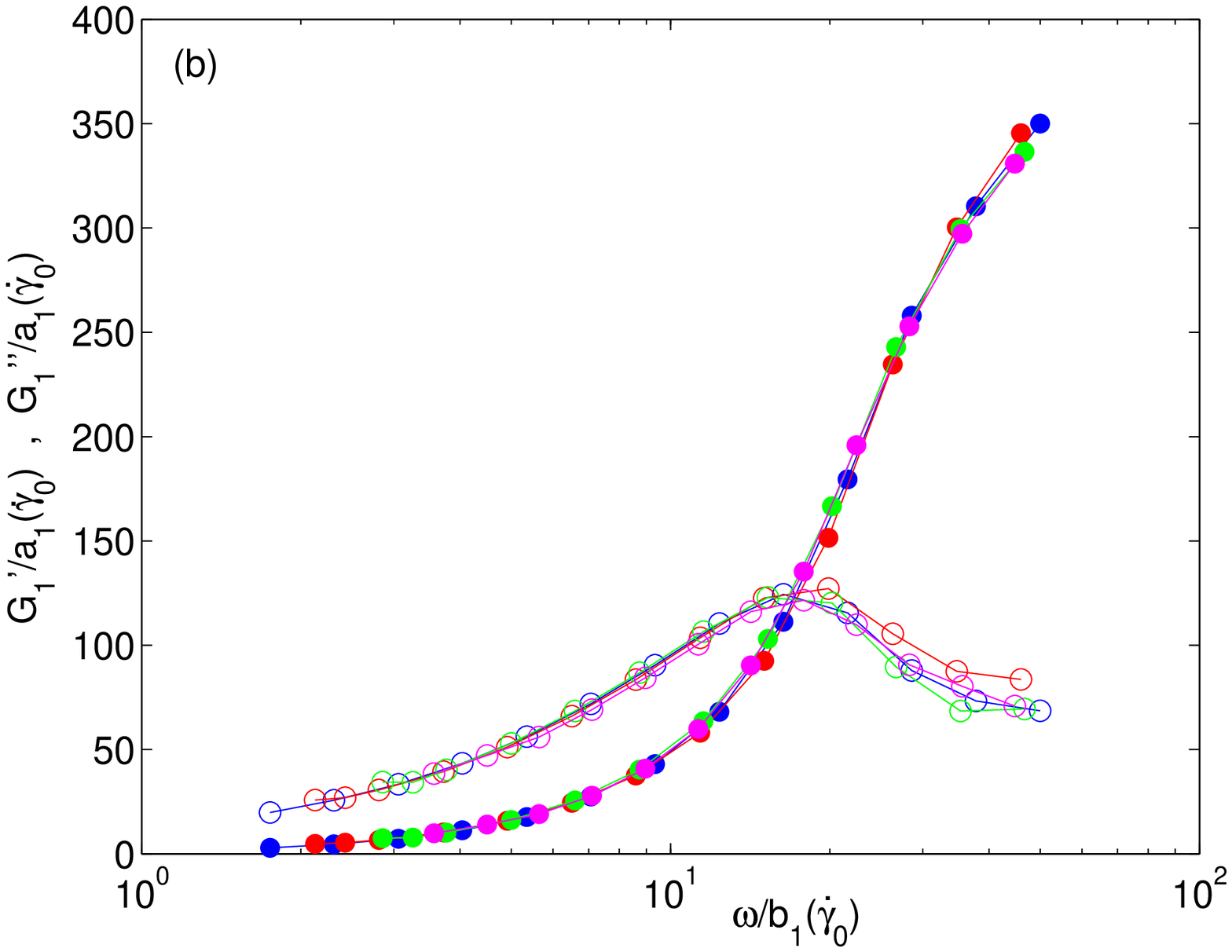}
\caption{\label{fig:srfs_pnipam1}(color online) (a) Plot of the first harmonic moduli $G'_1$ (solid circles) and $G''_1$ (open circles) as a function of the angular frequency $\omega$ from constant strain-rate frequency sweep measurements at strain-rate amplitudes $\dot{\gamma_0}=4.2$ $s^{-1}$ (blue), $2.1$ $s^{-1}$ (red), 
$1.2$ $s^{-1}$ (green), $0.9$ $s^{-1}$ (pink) using PNIPAM.\\(b) Plot of the scaled first harmonic moduli 
$G'_1/a_1(\dot{\gamma_0})$ (solid circles) and $G''_1/a_1(\dot{\gamma_0})$ (open circles) as a function of the scaled angular frequency $\omega/b_1(\dot{\gamma}_0)$ shifted onto a single master curve. See Table I for 
the values of $a_1(\dot{\gamma}_0)$ and $b_1(\dot{\gamma}_0)$.}
\end{figure}

In Fig. \ref{fig:srfs_pnipam1}(a), we plot the first harmonic moduli from SRFS tests with values of the strain-rate amplitude $\dot{\gamma_0}=\gamma_0\omega$ in the range [$0.9$ $s^{-1}$,$4.2$ $s^{-1}$], as a function of $\omega$ using PNIPAM. Note that the moduli crossover frequency is shifted in the direction of increasing $\omega$, with larger values of $\dot{\gamma_0}$ and the shapes of the moduli curves are similar. In Fig. \ref{fig:srfs_pnipam1}(b), we plot the rescaled moduli as a function of the rescaled angular frequency. The values of the shift factors $a_1(\dot{\gamma_0})$ and  $b_1(\dot{\gamma_0})$ are listed in Table I and the moduli curves for the highest value of $\dot{\gamma_0}$ are taken as the reference curves for the purpose of shifting. Note that although $G''_1$ is strictly positive (see Section III(E) for discussion), there is no restriction on the sign of the other moduli, which is why we plot all our moduli curves on a semilogarithmic scale for consistency. The plot confirms the validity of the SRFS procedure for linear viscoelastic moduli, as reported in Wyss, {\it et.al.}\cite{Wyss}. To ensure that the recorded torque signal was well within the capabilities of the machine, our experiments were not carried out at small strain-rate amplitudes, which is why we were unable to confirm the low-strain rate dependence of $b(\dot{\gamma_0})$, asymptotically expected to approach unity\cite{Wyss}. A discussion regarding the high-frequency response of the moduli may be found at the end of this Section.

\begin{figure}
\includegraphics[height=1.8in]{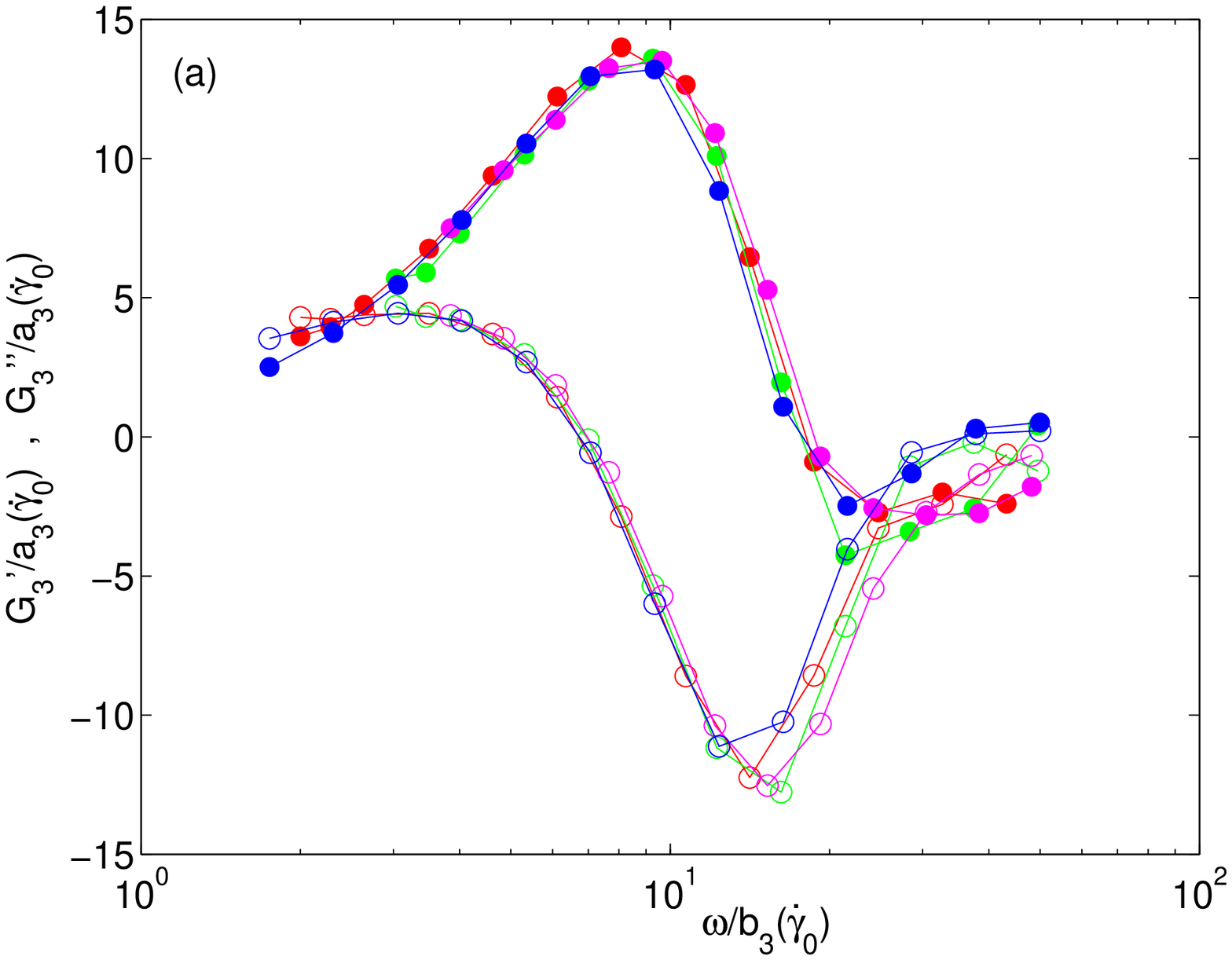}
\includegraphics[height=1.8in]{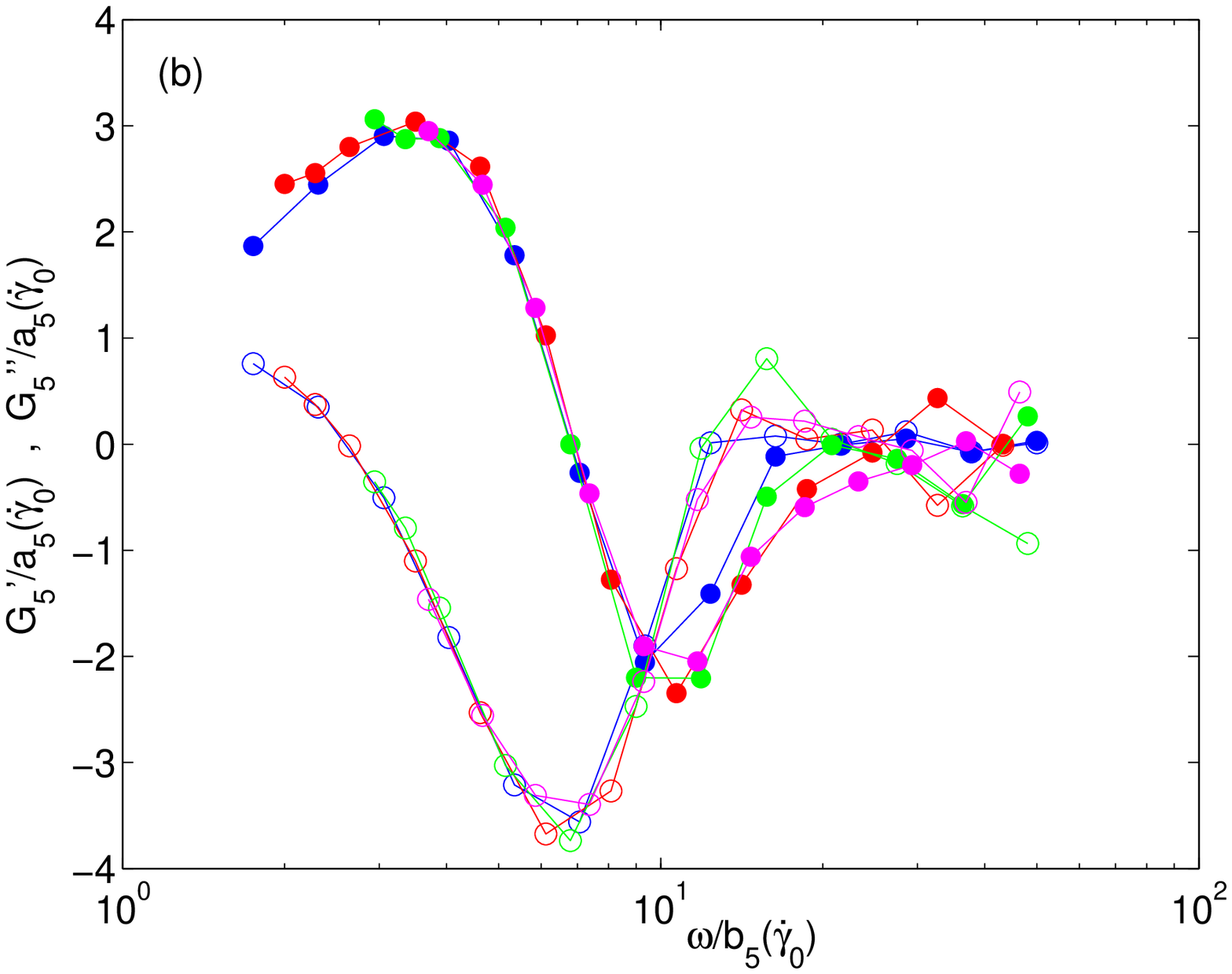}
\includegraphics[height=1.8in]{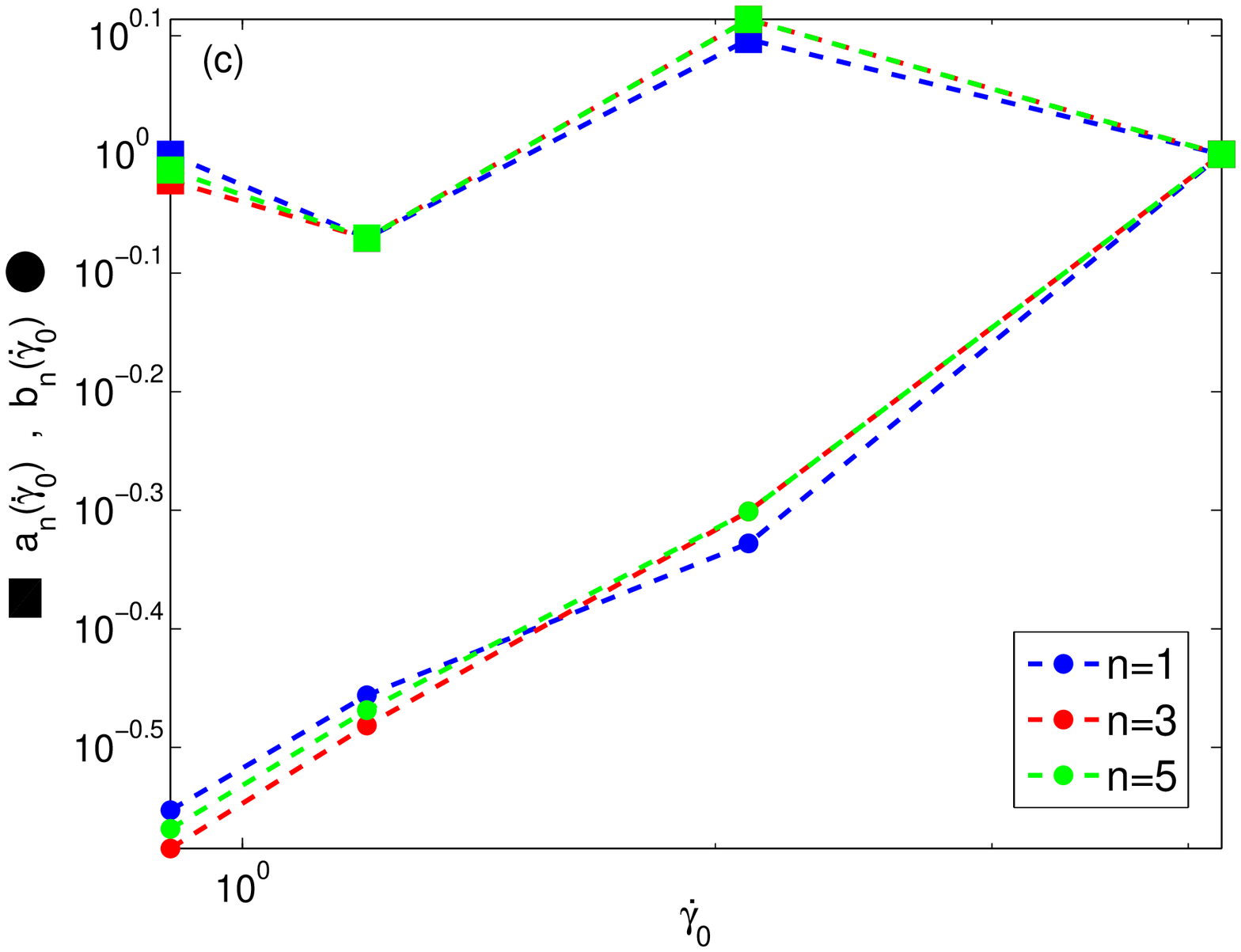}
\caption{\label{fig:srfs_pnipam2}(color online) (a) Plot of the scaled third harmonic moduli 
$G'_3/a_3(\dot{\gamma}_0)$ (solid circles) and $G''_3/a_3(\dot{\gamma}_0)$ (open circles) as a function of the scaled angular frequency $\omega/b_3(\dot{\gamma}_0)$ shifted onto a single master curve from constant strain-rate frequency sweep measurements at strain-rate amplitudes $\dot{\gamma_0}=4.2$ $s^{-1}$ (blue), $2.1$ $s^{-1}$ (red), $1.2$ $s^{-1}$ (green), $0.9$ $s^{-1}$ (pink) using PNIPAM. See Table I for the values of $a_3(\dot{\gamma}_0)$ and $b_3(\dot{\gamma}_0)$.\\(b) Corresponding plot of $G'_5/a_5(\dot{\gamma}_0)$ and $G''_5/a_5(\dot{\gamma}_0)$ as a function of $\omega/b_5(\dot{\gamma}_0)$. See Table I for the values of $a_5(\dot{\gamma}_0)$ and $b_5(\dot{\gamma}_0)$.\\(c) Plot of the shift factors $a_n(\dot{\gamma_0})$ (solid squares) and $b_n(\dot{\gamma_0})$ (solid circles) as a function of the strain-rate amplitude $\dot{\gamma_0}$ for the moduli $G'_n$ and $G''_n$ [$n=1$ (blue), $3$ (red), $5$ (green)].}
\end{figure}

Our results are, however, {\it not} restricted to the linear viscoelastic regime, and in Figs. \ref{fig:srfs_pnipam2}(a),(b), we show the corresponding rescaled third and fifth harmonic moduli as a function of the rescaled angular frequency from the same test as reported in Fig. \ref{fig:srfs_pnipam1}. The values of the shift factors are listed in Table I. Note the striking agreement between the values of the vertical shift factors ($a_3(\dot{\gamma_0})$, $a_5(\dot{\gamma_0})$) with $a_1(\dot{\gamma_0})$ and the horizontal shift factors ($b_3(\dot{\gamma_0})$, $b_5(\dot{\gamma_0})$) with $b_1(\dot{\gamma_0})$. In Fig. \ref{fig:srfs_pnipam2}(c), we plot the shift factors as a function of $\dot{\gamma_0}$. The shift factors $a_n(\dot{\gamma_0})$ ($n=1,3,5$) were found to be of order unity, while the shift factors $b_n(\dot{\gamma_0})$ ($n=1,3,5$) showed a power-law dependence on $\dot{\gamma_0}$ ($b_n(\dot{\gamma_0})\propto\dot{\gamma_0}^\nu$), with an exponent $\nu=0.89\pm0.01$. The shift factor $b_1(\dot{\gamma_0})$ is postulated\cite{Wyss} to depend inversely on the structural relaxation time, viz. $b_1(\dot{\gamma_0})\propto1/\tau(\dot{\gamma_0})$, and our results for $b_n(\dot{\gamma_0})$ ($n=1,3,5$) are in agreement with the exponent $\nu\approx0.9$ reported in Wyss, {\it et.al.}\cite{Wyss}. Physically, this suggests that soft solids have internal structures that relax faster in the presence of an imposed strain-rate (the inverse of which essentially ``sets" the structural relaxation time of the material).

\begin{figure}
\includegraphics[height=1.8in]{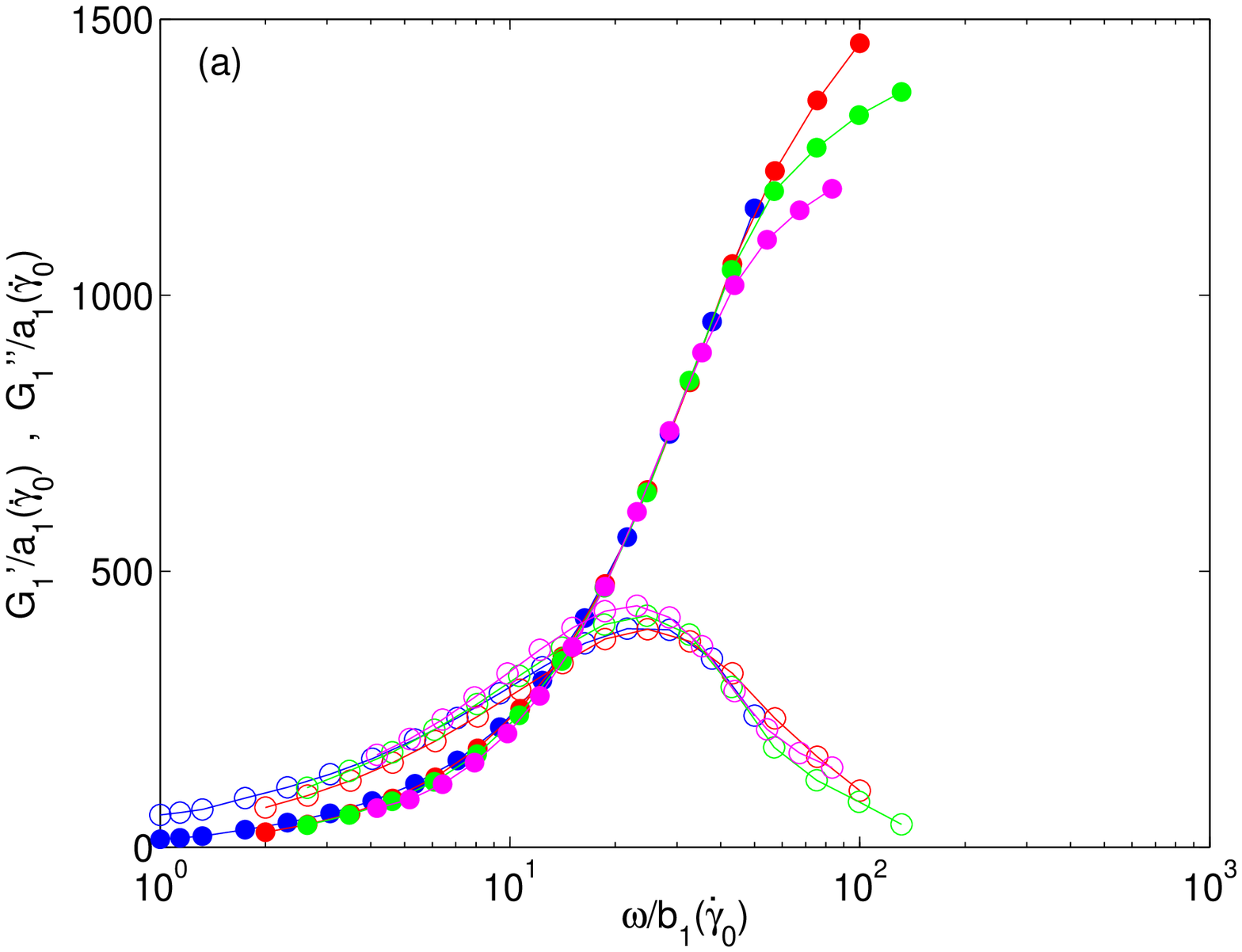}
\includegraphics[height=1.8in]{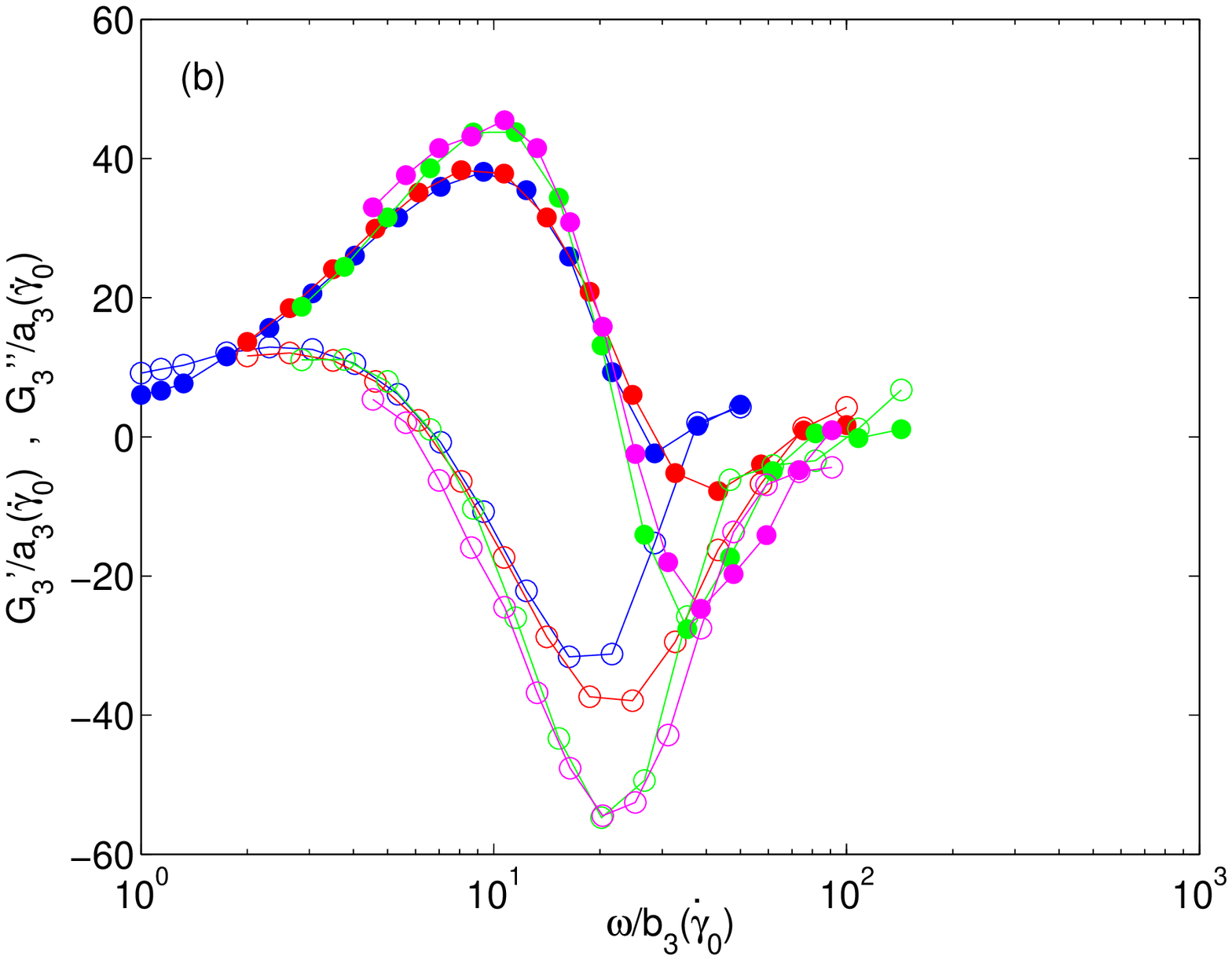}
\includegraphics[height=1.8in]{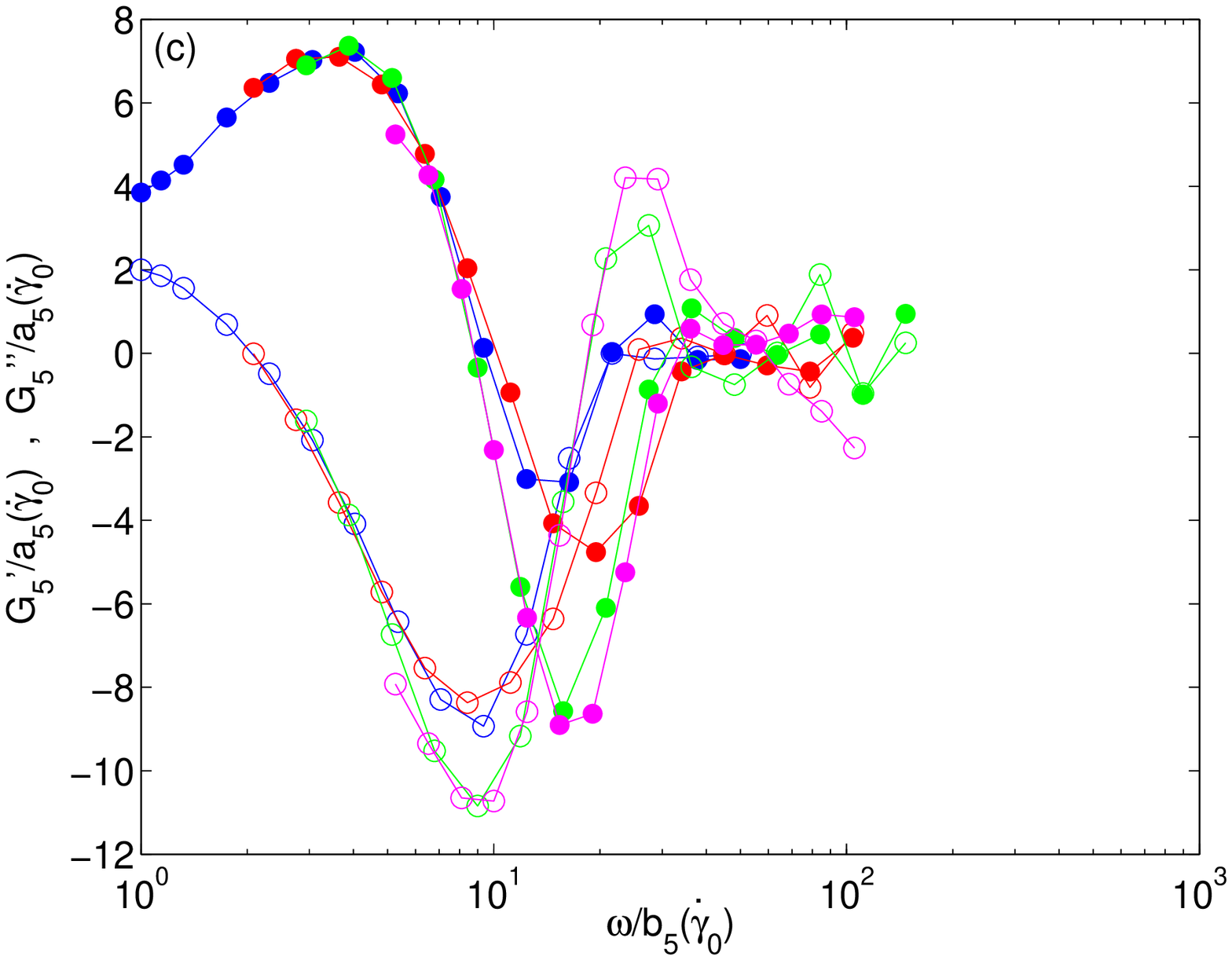}
\caption{\label{fig:srfs_xg}(color online) (a) Plot of the scaled first harmonic moduli 
$G'_1/a_1(\dot{\gamma}_0)$ (solid circles) and $G''_1/a_1(\dot{\gamma}_0)$ (open circles) as a function of 
the scaled angular frequency $\omega/b_1(\dot{\gamma}_0)$ shifted onto a single master curve from constant strain-rate frequency sweep measurements at strain-rate amplitudes $\dot{\gamma_0}=4.2$ $s^{-1}$ (blue), $2.1$ $s^{-1}$ (red), $1.2$ $s^{-1}$ (green), $0.3$ $s^{-1}$ (pink) using Xanthan gum. See Table I for the values of $a_1(\dot{\gamma}_0)$ and $b_1(\dot{\gamma}_0)$.\\(b) Corresponding plot of $G'_3/a_3(\dot{\gamma}_0)$ and $G''_3/a_3(\dot{\gamma}_0)$ as a function of $\omega/b_3(\dot{\gamma}_0)$. See Table I for the values of $a_3(\dot{\gamma}_0)$ and $b_3(\dot{\gamma}_0)$.\\(c) Corresponding plot of $G'_5/a_5(\dot{\gamma}_0)$ and $G''_5/a_5(\dot{\gamma}_0)$ as a function of $\omega/b_5(\dot{\gamma}_0)$. See Table I for the values of $a_5(\dot{\gamma}_0)$ and $b_5(\dot{\gamma}_0)$.} 
\end{figure}

Analogous master curves from tests using Xanthan gum are shown in Fig. \ref{fig:srfs_xg} for the first, third and the fifth harmonic moduli. The shift factors for the higher harmonic moduli were again found to agree (see Table I) with the shift factors for the first harmonic moduli. The horizontal shift factors $b_n(\dot{\gamma_0})$ ($n=1,3,5$) showed a power-law dependence on $\dot{\gamma_0}$ with an exponent $\nu=0.89\pm0.05$. Results from our Brycreem tests are not shown in this work, apart from the shift factors listed in Table I, and serve to confirm our primary result. In the case of Brycreem, the power-law dependence of $b_n(\dot{\gamma_0})$ on $\dot{\gamma_0}$ followed, with $\nu=0.99\pm0.11$.
\begin{table}
\caption{Vertical ($a_n(\dot{\gamma_0})$) and horizontal ($b_n(\dot{\gamma_0})$) shift factors for the harmonic moduli $G_n'$ and $G_n''$ ($n=1,3,5$) for different strain-rate amplitudes $\dot{\gamma_0}$ using PNIPAM, Xanthan gum and Brylcreem.}
\begin{tabular}{|c|c|c|c|c|c|c|c|}
\hline\hline
Material & $\dot{\gamma_0}$ [$s^{-1}$] & $a_1$ & $b_1$ & $a_3$ & $b_3$ & $a_5$ & $b_5$ \\ \hline
\multirow{4}{*}{PNIPAM} & $0.9$ & $1$ & $0.28$ & $0.95$ & $0.26$ & $0.97$ & $0.27$\\
 & $1.2$ & $0.85$ & $0.35$ & $0.85$ & $0.33$ & $0.85$ & $0.34$ \\
 & $2.1$ & $1.25$ & $0.47$ & $1.3$ & $0.5$ & $1.3$ & $0.5$ \\
 & $4.2$ & $1$ & $1$ & $1$ & $1$ & $1$ & $1$  \\ \hline
\multirow{4}{*}{Xanthan gum} & $0.3$ & $1.1$ & $0.12$ & $0.85$ & $0.095$ & $1$ & $0.095$ \\
 & $1.2$ & $1.1$ & $0.38$ & $0.9$ & $0.35$ & $1$ & $0.34$ \\
 & $2.1$ & $1.1$ & $0.5$ & $0.8$ & $0.5$ & $0.79$ & $0.48$ \\
 & $4.2$ & $1$ & $1$ & $1$ & $1$ & $1$ & $1$  \\ \hline
\multirow{4}{*}{Brylcreem} & $0.5$ & $0.83$ & $0.05$ & $0.85$ & $0.05$ & $0.91$ & $0.07$ \\
 & $1$ & $0.75$ & $0.09$ & $0.74$ & $0.1$ & $0.8$ & $0.12$ \\
 & $5$ & $0.83$ & $0.45$ & $0.83$ & $0.5$ & $0.83$ & $0.55$ \\
 & $10$ & $1$ & $1$ & $1$ & $1$ & $1$ & $1$ \\ \hline\hline
\end{tabular}
\label{table}
\end{table}

\begin{figure}
\includegraphics[height=1.8in]{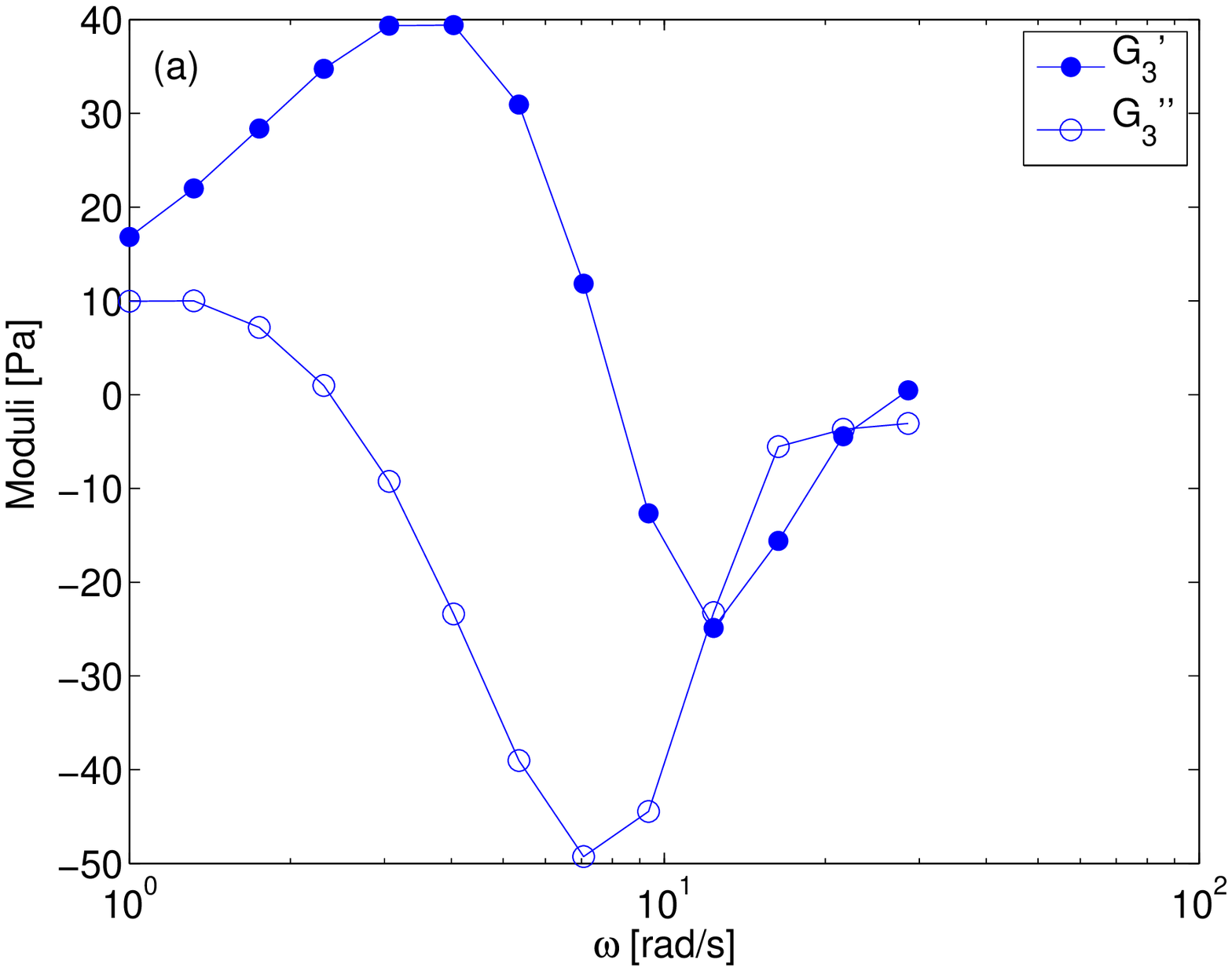}
\includegraphics[height=1.8in]{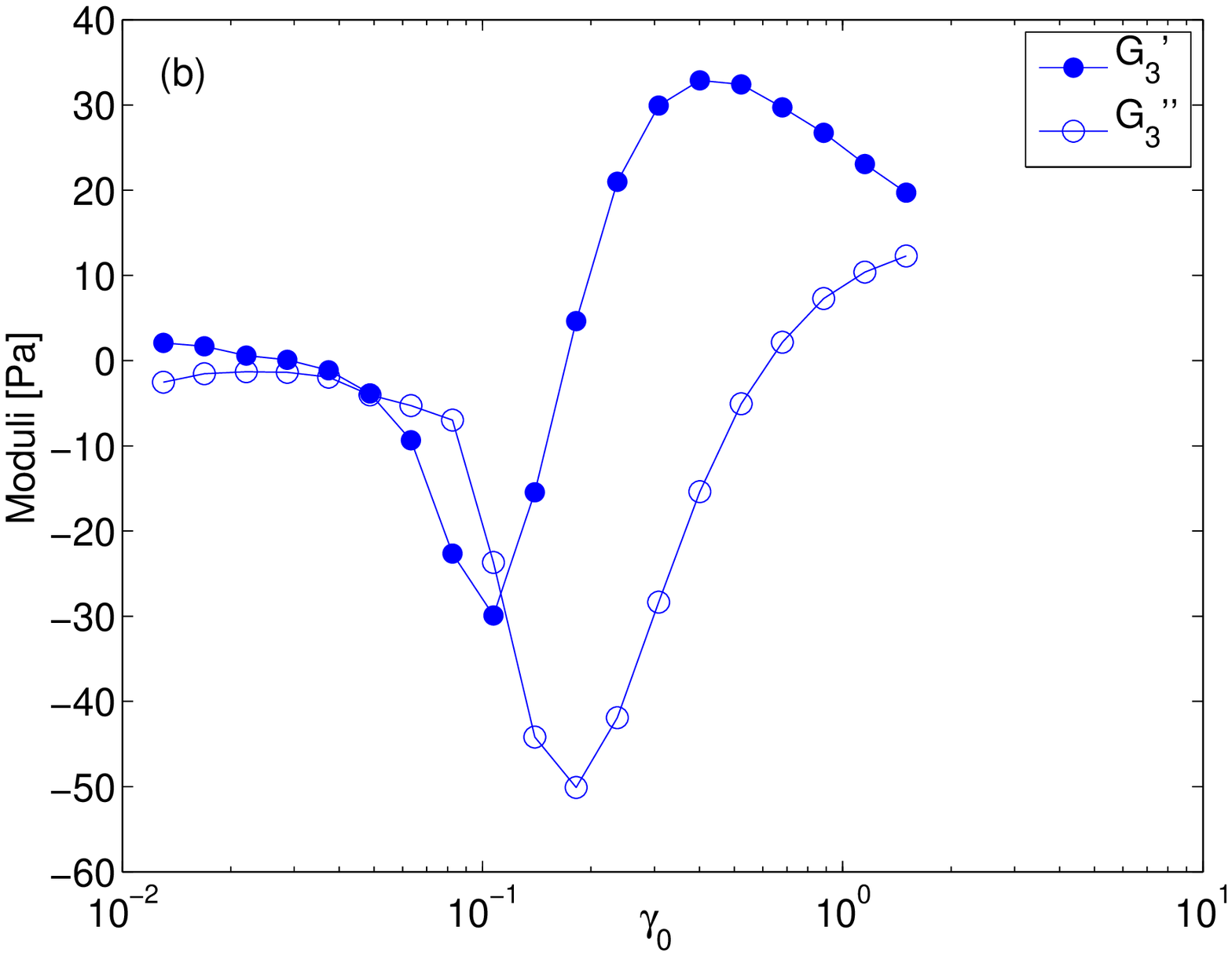}
\caption{\label{fig:mirror_xg}(a) Plot of the third harmonic moduli $G'_3$ (solid circles) and $G''_3$ (open circles) as a function of the angular frequency $\omega$ with constant strain-rate amplitude $\dot{\gamma_0}=1.2$ $s^{-1}$, using Xanthan gum.\\(b) Plot of the third harmonic moduli $G'_3$ (solid circles) and $G''_3$ (open circles) as a function of the strain amplitude $\gamma_0$ for angular frequency $\omega=5$ rad/s, using Xanthan gum.} 
\end{figure}

In Fig. \ref{fig:ampfreqsweep_pnipam}(b), the first harmonic moduli cross over at a frequency which is not easily accessible, however, the shape of the SRFS master curve in Fig. \ref{fig:srfs_pnipam1}(b), suggests that the low frequency behavior of the moduli is similar to that at high strain amplitudes (compare with Fig. \ref{fig:ampfreqsweep_pnipam}(a)). We confirm an analogous result for higher harmonic moduli: Figure \ref{fig:mirror_xg}(a) is a plot of the third harmonic moduli $G'_3$ and $G''_3$ as a function of $\omega$ from an SRFS test with $\dot{\gamma_0}=1.2$ $s^{-1}$ using Xanthan gum, while Fig. \ref{fig:mirror_xg}(b) is a plot from a varying strain amplitude test at $\omega=5$ rad/s. The plots are a remarkable demonstration of the `reversed' nature of the higher harmonic moduli as reported from SRFS tests and from a strain-amplitude sweep test. Similar results have been confirmed for the fifth harmonic moduli.

In Wyss, {\it et.al.}\cite{Wyss}, it is shown that at high frequencies, the curves for $G''_1$ for different strain-rate amplitudes can be rescaled onto a single master curve if a $\sqrt{\omega}$ component is subtracted, the high-frequency {\it linear} viscoelastic response being attributed to viscous flow along randomly oriented slip planes\cite{Liu}. However, in the angular frequency sweep for PNIPAM (see Fig. \ref{fig:ampfreqsweep_pnipam}(b)), the dashed-line fit to the high-frequency portion was found to be proportional to $\omega^{0.078}$. Clearly, our PNIPAM sample has not reached the asymptotic high-frequency regime at $\omega=100$ rad/s (the upper-limit for ARES-2000) where the scaling law applies. For Xanthan gum and Brylcreem, similar results were obtained, with a high-frequency scaling for $G''_1$ proportional to $\omega^{0.082}$ and $\omega^{0.26}$ respectively. Therefore, we did not attempt to correct for the observed lack of collapse of $G''_1$ onto a master curve in Figs. \ref{fig:srfs_pnipam1}(b) and \ref{fig:srfs_xg}(a) at high frequencies. Apart from this, an additional effect which must be accounted for is the inertia of the torque transducer and the variation of the torque transducer compliance due to vibrations transmitted through the frame of the rheometer, under high-frequency oscillations of the motor\cite{AFranck}. The ARES-$2000$ rheometer incorporates a hardware correction for these effects and there is no means by which the corrections may be implemented through an external software. The shift factors listed in Table I, however, remain unaffected.
\subsection{Surfaces of harmonic moduli}
Most LAOS studies to-date have been restricted to the study of stress-strain curves\cite{Philippoff,Tee,Gamota,Ewoldt} or the ratios of the harmonics of the stress amplitude spectrum\cite{Krieger,Davis,Ganeriwala}. In occasional studies\cite{Reimers}, the harmonic moduli have been calculated in a strain amplitude or an angular frequency sweep test. However, the complete form of the moduli is seen clearly in a surface plot, obtained from oscillation tests at different values of $\gamma_0$ and $\omega$. In an early study, Reimers and Dealy\cite{Reimers} used a sliding-plate rheometer and plotted surfaces of the third and the fifth harmonic moduli from LAOS studies with polystyrene. The plots in their study are qualitative at best and of low resolution.

\begin{figure}
\includegraphics[height=1.8in]{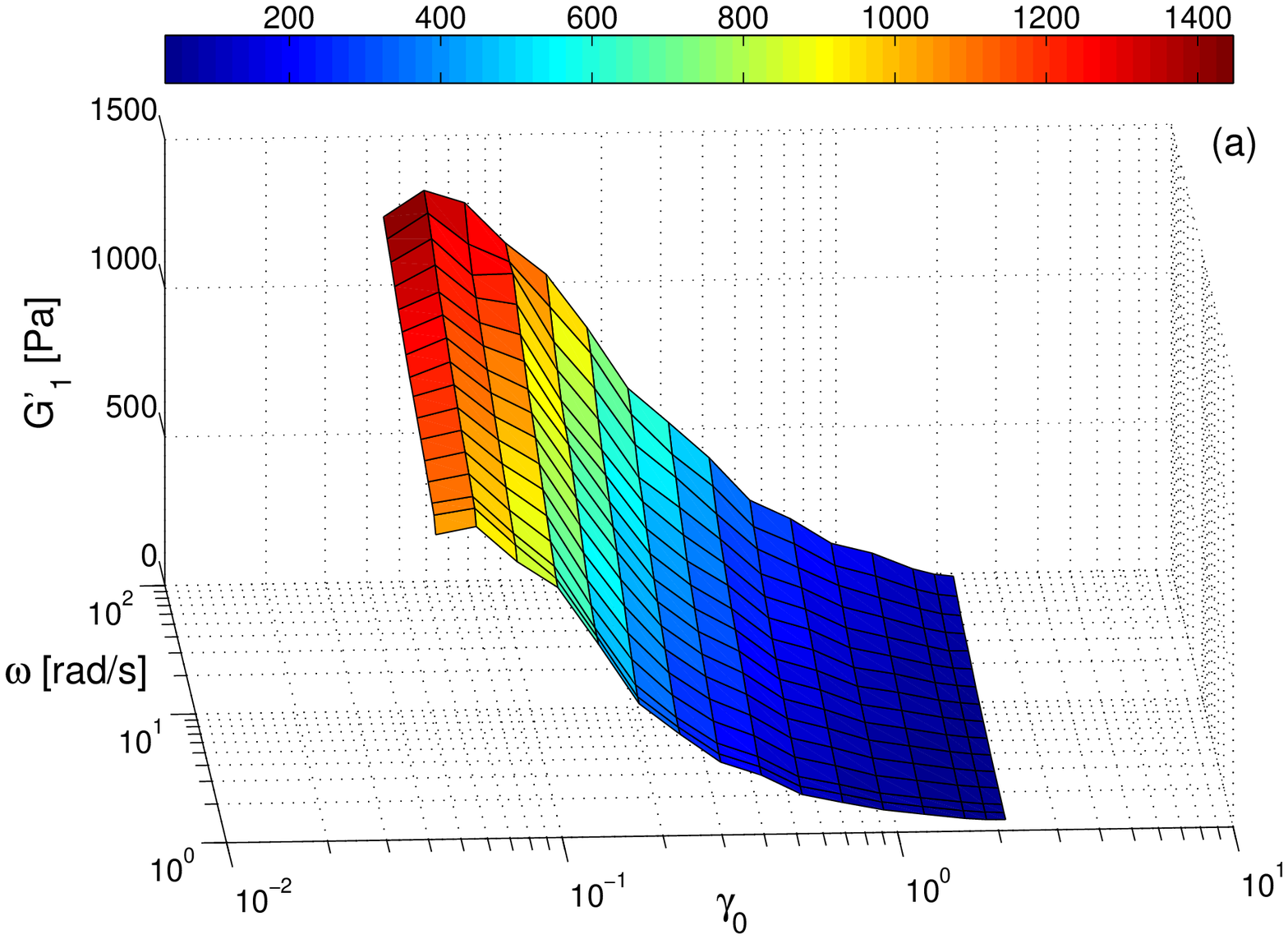}
\includegraphics[height=1.8in]{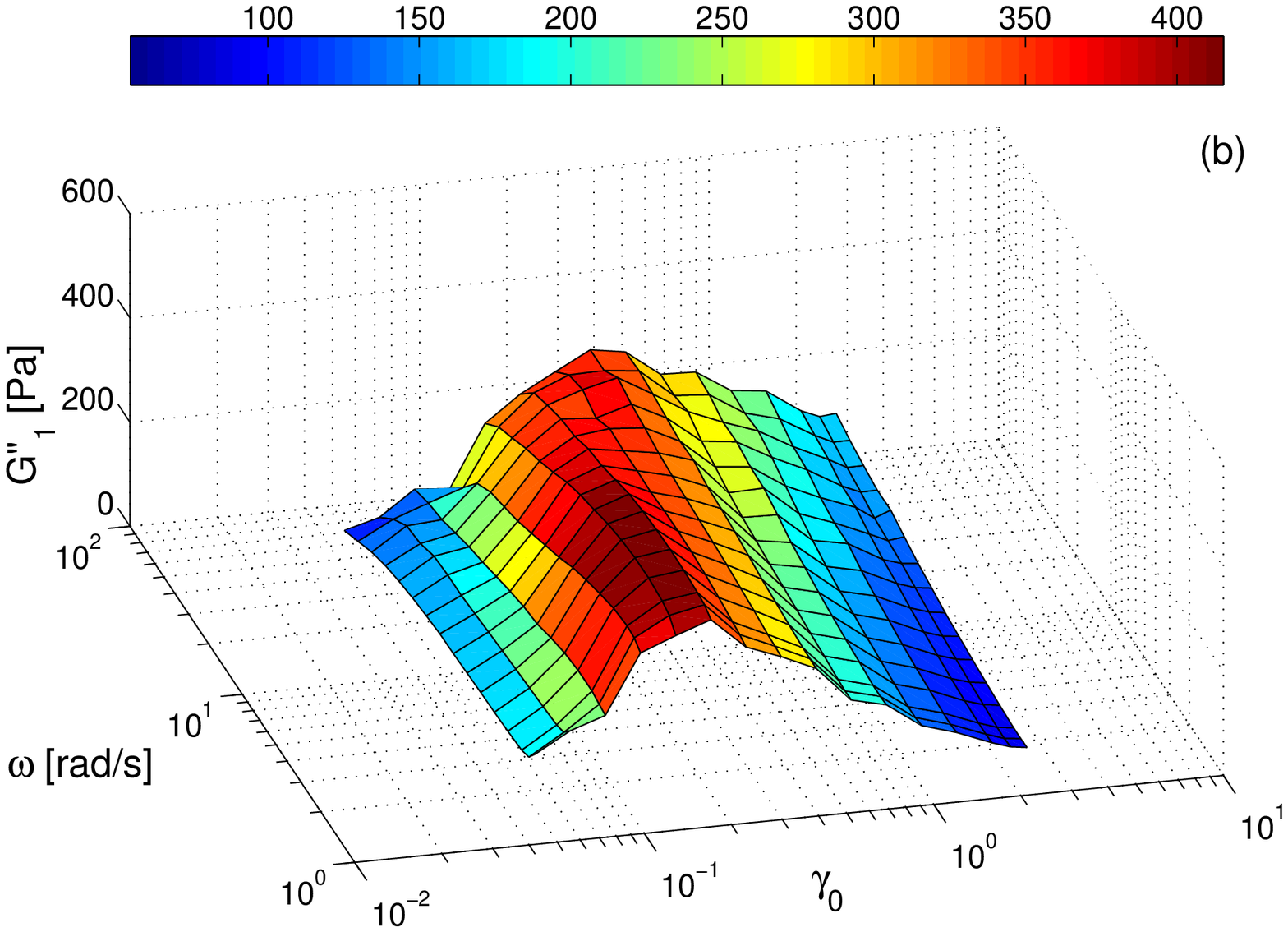}
\caption{\label{fig:surfg1_xg}(color online) (a) Surface plot of the first harmonic modulus $G'_1(\omega,\gamma_0)$ as a function of strain amplitude $\gamma_0$ and angular frequency $\omega$, using Xanthan gum. The color bar indicates magnitude of the modulus in units of Pascal.\\(b) Surface plot of the first harmonic modulus $G''_1(\omega,\gamma_0)$ as a function of strain amplitude $\gamma_0$ and angular frequency $\omega$, using Xanthan gum.} 
\end{figure}

\begin{figure}
\includegraphics[height=1.8in]{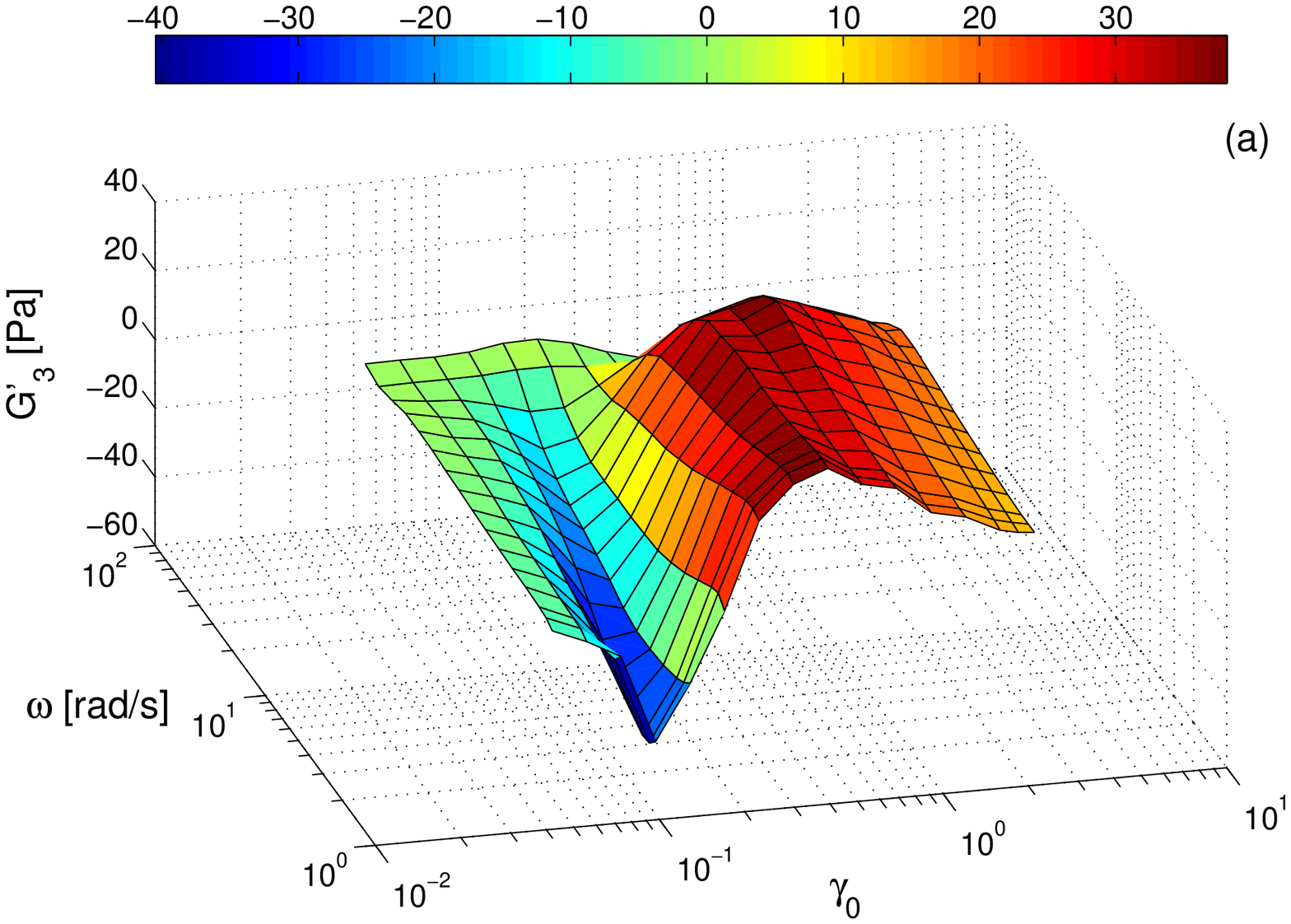}
\includegraphics[height=1.8in]{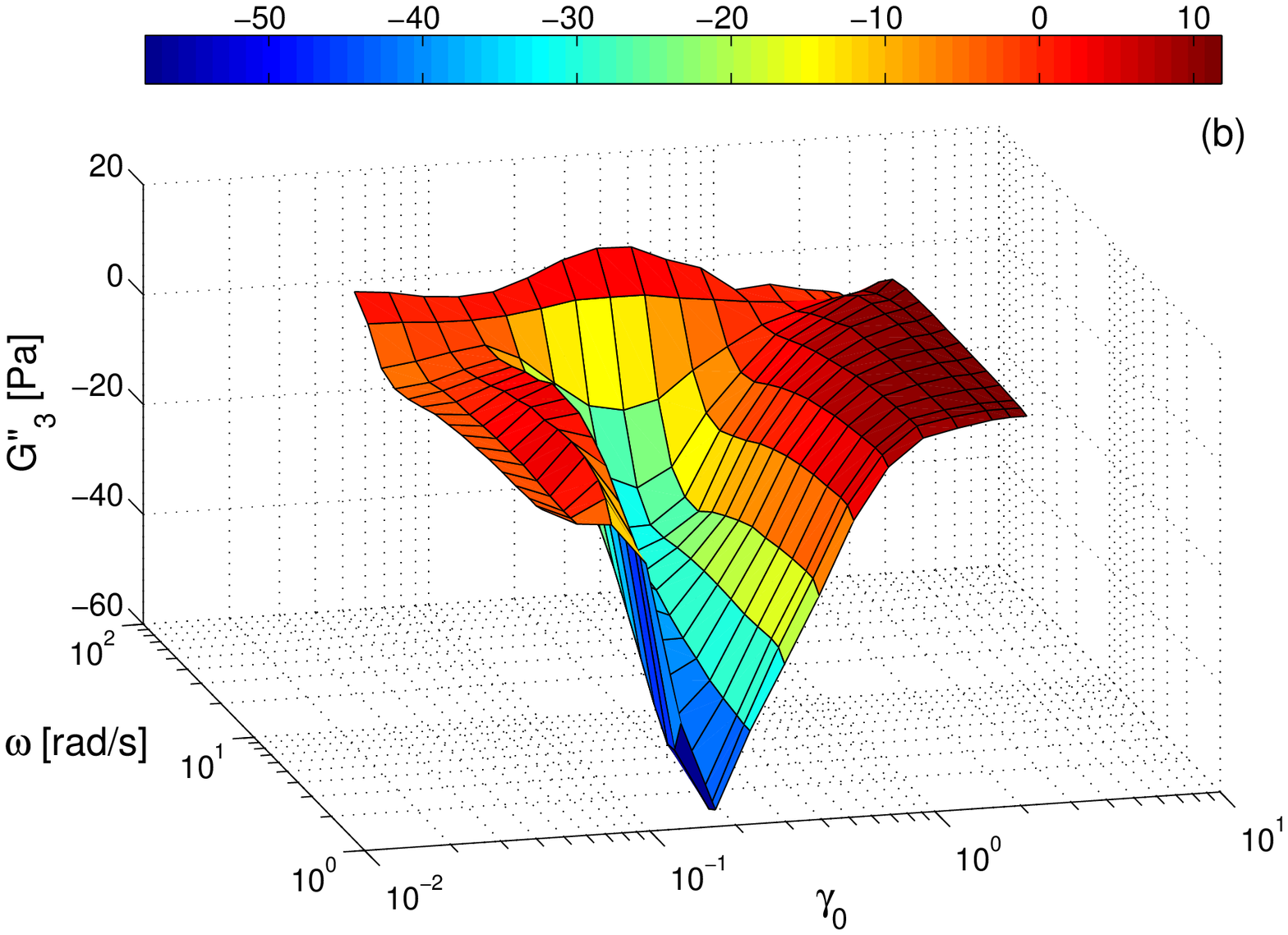}
\caption{\label{fig:surfg3_xg}(color online) (a) Surface plot of the third harmonic modulus $G'_3(\omega,\gamma_0)$ as a function of strain amplitude $\gamma_0$ and angular frequency $\omega$, using Xanthan gum. The color bar indicates magnitude of the modulus in units of Pascal.\\(b) Surface plot of the third harmonic modulus $G''_3(\omega,\gamma_0)$ as a function of strain amplitude $\gamma_0$ and angular frequency $\omega$, using Xanthan gum.} 
\end{figure}

\begin{figure}
\includegraphics[height=1.8in]{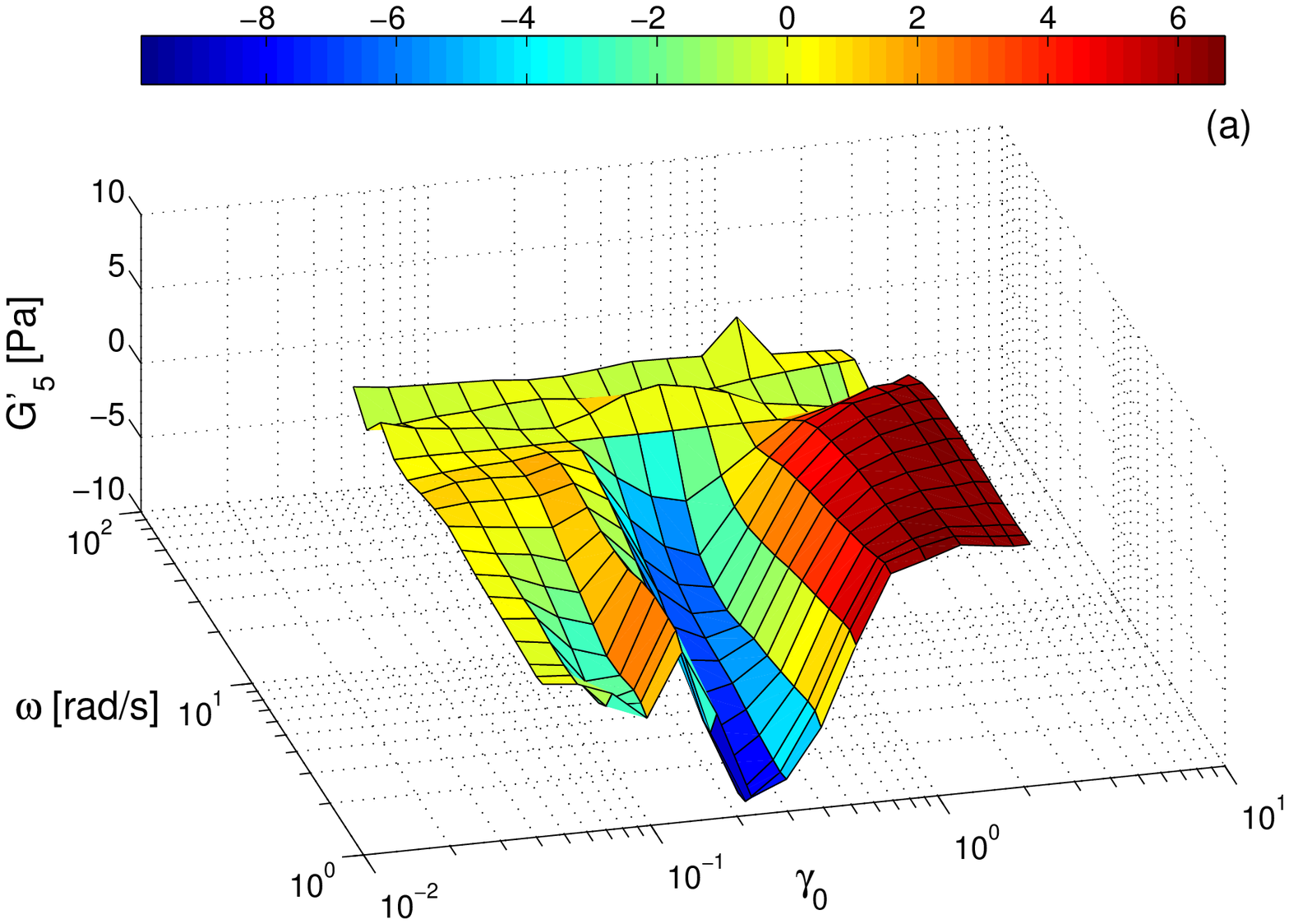}
\includegraphics[height=1.8in]{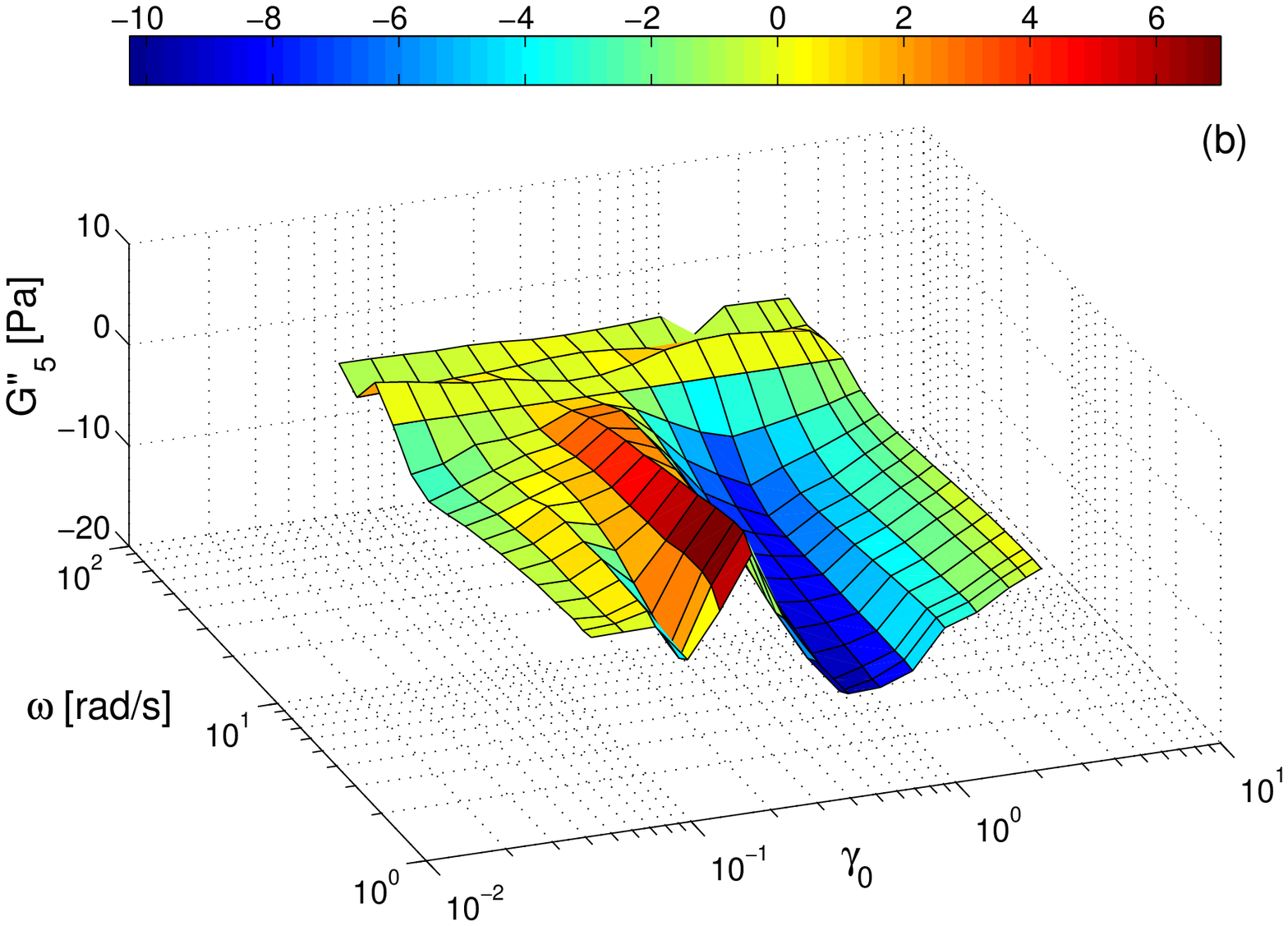}
\caption{\label{fig:surfg5_xg}(color online) (a) Surface plot of the fifth harmonic modulus $G'_5(\omega,\gamma_0)$ as a function of strain amplitude $\gamma_0$ and angular frequency $\omega$, using Xanthan gum. The color bar indicates magnitude of the modulus in units of Pascal.\\(b) Surface plot of the fifth harmonic modulus $G''_5(\omega,\gamma_0)$ as a function of strain amplitude $\gamma_0$ and angular frequency 
$\omega$, using Xanthan gum.} 
\end{figure}

In Figs. \ref{fig:surfg1_xg}, \ref{fig:surfg3_xg}, and \ref{fig:surfg5_xg}, we show surface plots of the harmonic moduli. These plots were obtained from $256$ independent oscillatory shear tests using Xanthan gum, with the values of $\gamma_0$ and $\omega$ spaced logarithmically on a $16\times16$ grid, the elements along one diagonal having a strain-rate amplitude $\dot{\gamma_0}=\gamma_0\omega=2.1$ $s^{-1}$ (a value chosen from Table I). Although the first harmonic loss modulus $G''_1$ is strictly positive (see Section III(E) below), the other moduli can be of any sign, which is why our surface plots are shown on a logarithmic scale in the $(\omega,\gamma_0)$ plane, and with a linear scale on the third axis. Note that the SRFS results in Wyss, {\it et.al.}\cite{Wyss} pertain to the study of curves of intersection of the hyperbolic sheets $\gamma_0\omega=\Gamma$ ($\Gamma$ are the chosen values of the strain-rate amplitudes, here $\Gamma=2.1$ $s^{-1}$) with the corresponding harmonic surfaces for $G'_1(\omega,\gamma_0)$ and $G''_1(\omega,\gamma_0)$.  Similarly, the intersection of the hyperbolic sheet $\Gamma=2.1$ $s^{-1}$ with the harmonic surfaces plotted in Figs. \ref{fig:surfg3_xg} and \ref{fig:surfg5_xg} would supply the curves marked with red circles in Figs. \ref{fig:srfs_xg}(b) and (c) respectively.

We observe that the surface for $G''_1$ has a local {\it maxima} at $(\omega,\gamma_0)=(2.31$ rad/s$,0.17)$, while the surfaces for $G'_3$, $G''_3$, $G'_5$, and $G''_5$ have local {\it minimas} at $(\omega,\gamma_0)=(1.32$ rad/s$,0.097)$,$(1.14$ rad/s$,0.17)$,$(1.32$ rad/s$,0.225)$, and $(1.14$ rad/s$,0.393)$ respectively. The moduli $G''_1$ and $G''_3$ share the same value of $\gamma_0$, while the pairs $(G'_3, G'_5)$ and $(G''_3,G''_5)$ share the same values of $\omega$. This interesting coupling between the moduli requires further investigation in surface plots with much higher resolution, and may be of interest to rheologists modelling soft glassy materials.
\subsection{Energy dissipation and dissipation rate}
The energy dissipated per unit volume per cycle of strain oscillation is $\epsilon(t;\omega,\gamma_0)\equiv\int_0^{2\pi/\omega}\sigma d\gamma$\cite{Onogi}. Ganeriwala and Rotz\cite{Ganeriwala} have shown that on substituting the one-dimensional Green-Rivlin constitutive equation\cite{greenrivlin} for the stress into this formula and assuming a sinusoidal strain one obtains $\epsilon=\pi\gamma_0^2G_1''(\omega,\gamma_0)$, an expression which is true for arbitrary values of $\gamma_0$. The second law of thermodynamics requires that $\epsilon\ge 0$, i.e. $G''_1$ is strictly positive, but places no restriction on the sign of the other moduli. It is plausible that the other harmonic moduli are involved in reversible exchanges of energy in LAOS.

\begin{figure}
\includegraphics[height=1.8in]{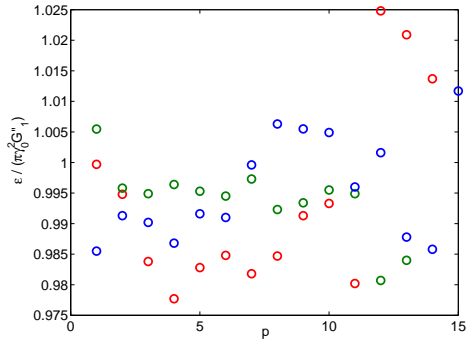}
\caption{\label{fig:epsilon}(color online) Plot of $\epsilon/(\pi\gamma_0^2G''_1)$ as a function of the number of the point $p$ (each point representing an independent test) from three different tests, a constant strain-rate frequency sweep test (red) at $\dot{\gamma_0}=0.9$ $s^{-1}$ using PNIPAM, a strain-amplitude sweep test (blue) at $\omega=9.35$ rad/s and a frequency sweep test (green) at $\gamma_0=0.52$, both using Xanthan gum.} 
\end{figure}

In Fig. \ref{fig:epsilon}, we plot the ratio $\epsilon/(\pi\gamma_0^2G''_1)$ ($\epsilon$ being the area bounded by the stress-strain curve\cite{green}), as a function of the number of the point $p$ (each point representing an independent test) for three different tests: A constant strain-rate frequency sweep test at $\dot{\gamma_0}=0.9$ $s^{-1}$ ($\omega$ in the range $[1,25]$ rad/s, $\gamma_0$ in the range $[0.036,0.9]$) using PNIPAM, a frequency sweep test at $\gamma_0=0.52$ ($\omega$ in the range $[1,25]$ rad/s) using Xanthan gum, and a strain-amplitude sweep test at $\omega=9.35$ rad/s ($\gamma_0$ in the range $[0.042,2.1]$) using Xanthan gum. In all three test configurations, we note that the points cluster about the expected value of $1$, with a maximum deviation of approximately $2\%$, the residual difference being attributed to experimental uncertainty in the measurement of $G''_1$.

\begin{figure}
\includegraphics[height=1.8in]{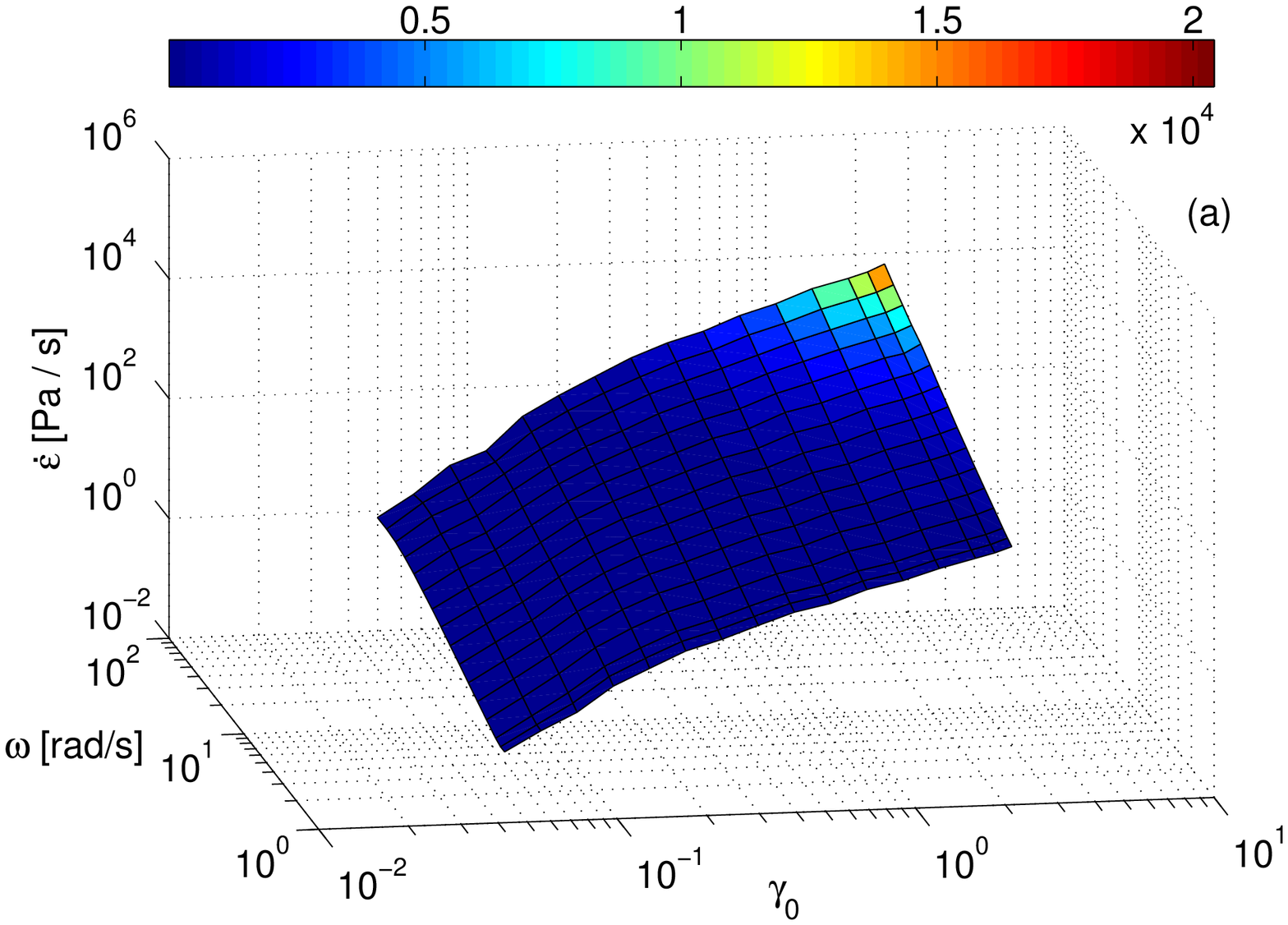}
\includegraphics[height=1.8in]{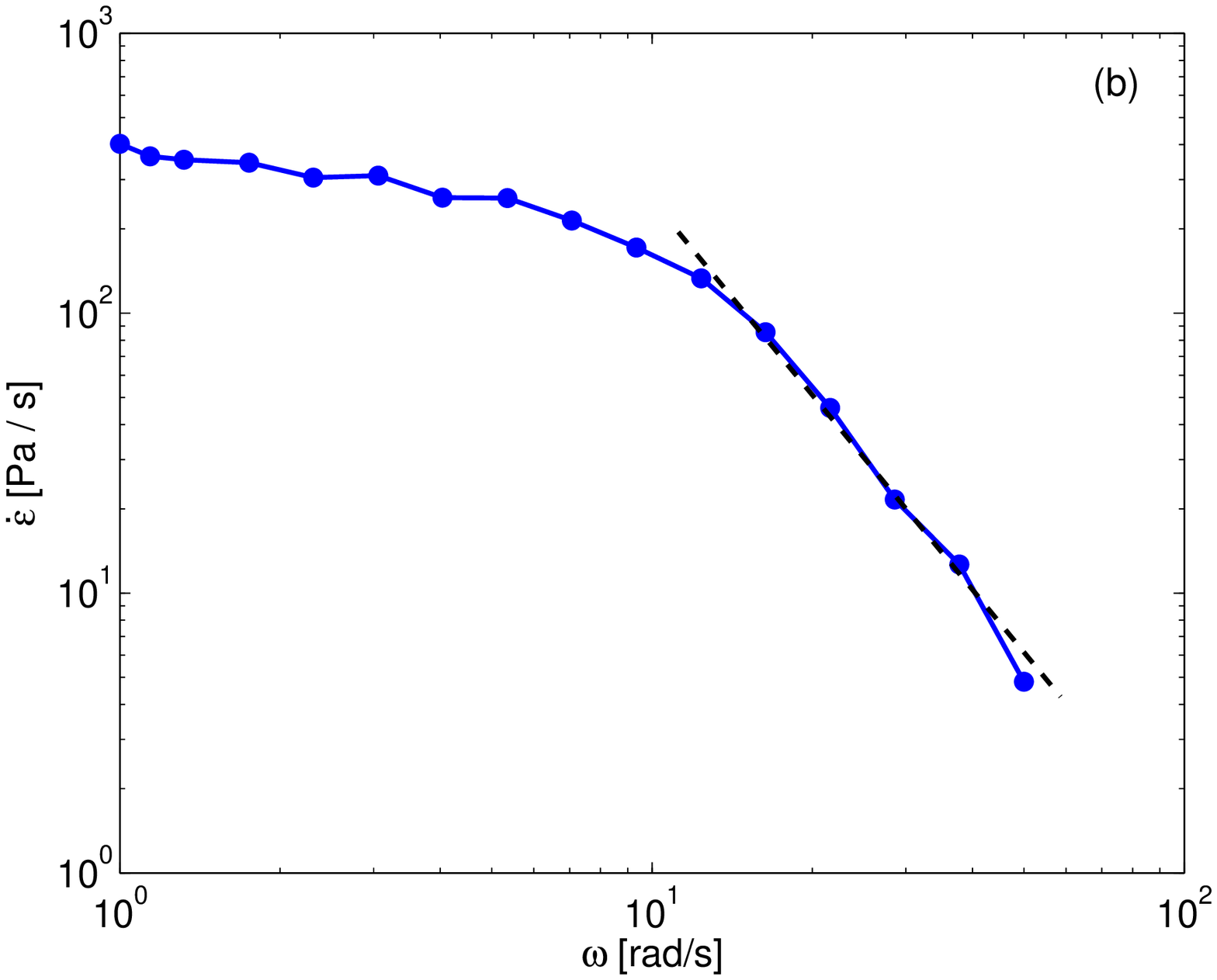}
\caption{\label{fig:edissrate_xg}(a) (color online) Surface plot of the energy-dissipation rate per unit volume 
$\dot{\epsilon}(t;\omega,\gamma_0)\equiv\omega\gamma_0^2G''_1(\omega,\gamma_0)/2$ as a function of the angular frequency $\omega$ and the strain amplitude $\gamma_0$, using Xanthan gum. The color bar indicates the magnitude of $\dot{\epsilon}$ in units of Pa/s.\\(b) Curve of intersection of the surface 
$\dot{\gamma_0}=\omega\gamma_0=2.1$ $s^{-1}$ with the surface plot in (a). The dashed line is proportional to $\omega^{-2.36}$.} 
\end{figure}

The energy-dissipation rate per unit volume in LAOS can be shown\cite{Ganeriwala} to equal $\dot{\epsilon}(t;\omega,\gamma_0)=\omega\gamma_0^2G_1''(\omega,\gamma_0)/2$. In Fig. \ref{fig:edissrate_xg}(a), we show the surface plot of the energy-dissipation rate per unit volume for Xanthan gum. It is interesting to note that the 
$\dot{\epsilon}$ surface grows monotonically with $\gamma_0$ and $\omega$, despite the material showing a solid-like response at high frequencies. On account of the logarithmic scaling used for the axes and the orientation of the surface in Fig. \ref{fig:edissrate_xg}(a), it may not be directly apparent that 
$\dot{\epsilon}$ is in fact a {\it decreasing} function at large $\omega$ along the curve of intersection of the surface $\dot{\gamma_0}=\omega\gamma_0=2.1$ $s^{-1}$ with the surface for $\dot{\epsilon}$, as is shown in Fig. \ref{fig:edissrate_xg}(b). We may rewrite the expression for the energy dissipation rate per unit volume as $\dot{\epsilon}=\dot{\gamma_0}^2G''_1/(2\omega)$. In linear viscoelasticity, $G''_1\propto\omega$ at small $\omega$\cite{Ferry}, implying that $\dot{\epsilon}$ is constant along the SRFS curve (with constant 
$\dot{\gamma_0}$) at small $\omega$, which accords with Fig. \ref{fig:edissrate_xg}(b) for 
$\dot{\gamma_0}=2.1$ $s^{-1}$. The dashed-line fit to the high-frequency (corresponding to {\it low} strain amplitude) portion of the $\dot{\epsilon}$ curve was found to be proportional to $\omega^{-2.36}$. For comparison, in linear viscoelasticity, the Maxwell model (see Ref. \cite{Ferry}, p. 57) predicts $\dot{\epsilon}\propto1/\omega^2$ along the SRFS curve at high frequencies.
\section{Summary}
To conclude, we have presented results from a systematic experimental study of soft solids under LAOS with special attention to SRFS measurements. Our results show the general applicability of the SRFS result for soft solids, specifically, that the SRFS curves for higher harmonic moduli can be superimposed onto master curves, with the same shift factors as for the linear viscoelastic moduli. We have also shown surface plots of the moduli and the energy-dissipation rate per unit volume in LAOS. We have shown that the energy dissipated per unit volume in oscillatory shear is governed by the first harmonic loss modulus alone. We hope that the results in this paper motivate further studies of LAOS in polymer melts, electro- and magnetorheological fluids, among other complex materials.
\begin{acknowledgments}
We thank Aloyse Franck, Allen Glasman (TA Instruments, United States) for technical support, Tejas Kalelkar, Chandan Dasgupta, and Sriram Ramaswamy for discussions.
\end{acknowledgments}


\begin{thebibliography}{11}
\bibitem{cp} By design, in a cone-plate measuring system the stresses and strains or strain rates are uniform 
across the plate gap.
\bibitem{nyquist} The sampling rate is well beyond the Nyquist rate of our experiments, aliasing effects are therefore neglected.
\bibitem{autorange} The torque transducer Autoranging feature was disabled, to prevent changes in torque scaling during the course of the experiments.
\bibitem{ampspec} For the real-valued time series $h(t)$ with Fourier coefficients $H(\omega)$, the amplitude spectrum is defined as $2|H(\omega)|$. The sign of the moduli $G'_n$, $G''_n$ are determined by the phase angles $\Phi_n$.
\bibitem{sinc} The sinc function is defined as $sinc(t)=1$ for $t=0$, $sin(\pi t)/(\pi t)$ for other values of $t$.
\bibitem{nonlinmon} The highest value of the ratio of the third to the first harmonic stress amplitude $I(3\omega)/I(\omega)\equiv\sigma_3/\sigma_1$ in our experiments was found to equal $0.22$, consistent with machine limitations.
\bibitem{greenrivlin} The odd-order Green-Rivlin equation in one dimension is $\sigma(t)=\int^t_{-\infty}K_1(t-t_1)\dot{\gamma}(t_1)dt_1+\int^t_{-\infty}\int^t_{-\infty}\int^t_{-\infty}K_3(t-t_1,t-t_2,t-t_3)\dot{\gamma}(t_1)\dot{\gamma}(t_2)\dot{\gamma}(t_3)dt_1dt_2dt_3+...$, where $K_1,K_3,...$ are stress-relaxation moduli (see Ref. \cite{Ferry} for the definition).
\bibitem{green} The area bounded by the stress-strain curve may be found by applying Green's theorem, viz. $\int_C (fdx + gdy)=\iint_D (\partial_x g-\partial_y f) dxdy$ where the plane region $D$ is bounded by the simple, closed curve $C$ with $f(x,y)$, $g(x,y)$ defined on an open region containing $D$ and having continuous partial derivatives there. For  $f=0$, $g=x$, the theorem reduces to $\int_C xdy=\iint_D dxdy$.
\end{thebibliography}
\end{document}